\documentclass[aps,onecolumn,amsmath,showpacs,superscriptaddress,floatfix,nofootinbib,nopreprintnumbers]{revtex4-2}

\usepackage{verbatim}
\usepackage[T1]{fontenc}
\usepackage[utf8]{inputenc}
\usepackage[american]{babel}
\usepackage{epsfig}
\usepackage{graphicx}
\usepackage{booktabs}
\usepackage{multirow}
\usepackage{dcolumn}
\usepackage{amsmath}
\usepackage{amssymb}
\usepackage{mathtools}
\usepackage{amsfonts}
\usepackage{amssymb}
\usepackage{epstopdf}
\usepackage{bm}
\usepackage{siunitx}
\usepackage{braket}
\usepackage{enumitem}
\usepackage{soul}
\usepackage[table]{xcolor}
\usepackage{color}
\usepackage{transparent}
\usepackage{pifont}
\usepackage[normalem]{ulem}

\definecolor{navyblue}{rgb}{0.0, 0.0, 0.5}
\definecolor{royalblue}{rgb}{0.25, 0.41, 0.88}
\definecolor{cadmiumgreen}{rgb}{0.0, 0.42, 0.24}
\definecolor{blue-violet}{rgb}{0.54, 0.17, 0.89}
\definecolor{darkviolet}{rgb}{0.58, 0.0, 0.83}
\definecolor{orange(colorwheel)}{rgb}{1.0, 0.5, 0.0}

\usepackage[bookmarks=true,bookmarksnumbered=true,colorlinks=true,urlcolor=royalblue, citecolor=blue,linkcolor=royalblue,breaklinks=true]{hyperref}
\hypersetup{
    colorlinks=true, 
    linkcolor=royalblue, 
    citecolor=blue}

\renewcommand\[{\left[}

\usepackage{booktabs}
\usepackage{multirow}
\usepackage{dcolumn}
\usepackage{colortbl}

\definecolor{magenta(process)}{rgb}{1.0, 0.0, 0.56}

\definecolor{darkspringgreen}{rgb}{0.09, 0.45, 0.27}

\definecolor{royalblue(web)}{rgb}{0.25, 0.41, 0.88}

\begin{document}

\title{Earth tomography with supernova neutrinos at future neutrino detectors}

\author{Rasmi Hajjar}
\email{rasmi.hajjar@ific.uv.es}

\affiliation{Instituto de F\'{i}sica Corpuscular (IFIC), University of Valencia-CSIC, Parc Cient\'{i}fic UV, C/ Catedr\'{a}tico Jos\'{e} Beltr\'{a}n 2, E-46980 Paterna, Spain}
\affiliation{Scuola Superiore Meridionale, Largo San Marcellino 10, 80138 Napoli, Italy}

\author{Olga Mena}
\email{omena@ific.uv.es}

\author{Sergio Palomares-Ruiz}
\email{sergiopr@ific.uv.es}

\affiliation{Instituto de F\'{i}sica Corpuscular (IFIC), University of Valencia-CSIC, Parc Cient\'{i}fic UV, C/ Catedr\'{a}tico Jos\'{e} Beltr\'{a}n 2, E-46980 Paterna, Spain}

\begin{abstract}
Earth neutrino tomography is a realistic possibility with current and future neutrino detectors, complementary to geophysics methods. The two main approaches are based on either partial absorption of the neutrino flux as it propagates through the Earth (at energies about a few TeV) or on coherent Earth matter effects affecting the neutrino oscillations pattern (at energies below a few tens of GeV). In this work, we consider the latter approach focusing on supernova neutrinos with tens of MeV. Whereas at GeV energies, Earth matter effects are driven by the atmospheric mass-squared difference, at energies below $\sim 100$~MeV, it is the solar mass-squared difference what controls them. Unlike solar neutrinos, which suffer from significant weakening of the contribution to the oscillatory effect from remote structures due to the neutrino energy reconstruction capabilities of detectors, supernova neutrinos can have higher energies and thus, can better probe the Earth's interior. We shall revisit this possibility, using the most recent neutrino oscillation parameters and up-to-date  supernova neutrino spectra. The capabilities of future neutrino detectors, such as DUNE, Hyper-Kamiokande and JUNO are presented, including the impact of the energy resolution and other factors. Assuming a supernova burst at 10~kpc, we show that the average Earth’s core density could be determined within $\lesssim 10\%$ at $1\sigma$ confidence level, being Hyper-Kamiokande, with its largest mass, the most promising detector to achieve this goal.
\end{abstract}

\maketitle

\section{Introduction} 
\label{sec:intro} 

The Earth is a nearly spherically symmetric body, with just a few major internal discontinuities. Its state at every evolutionary stage is characterized by the changes in time of its temperature, pressure, density and composition, which determine the interaction among its different components as a planetary system~\cite{Condie20161}. Determining the internal structure of the Earth becomes pivotal to understand its thermal history, which is the major factor driving Earth's evolution. In particular, it helps to estimate when the molten outer core was formed, which allowed the planet to develop a global magnetic field, critical for life on Earth. 

Geodesic, geomagnetic and geodynamical data, and theoretical and experimental results on the behavior of matter at high temperatures and pressures, are of great importance in the study of the interior of the Earth. Nonetheless, our knowledge about the internal structure of the Earth comes mainly from seismological data. Soon after the first detection, in 1889, of seismic waves produced by a distant earthquake, the idea of a global seismological network took form, during the early years of the twentieth century~\cite{Agnew20023, Dziewonski20151}. Since then, developments in theory, instrumentation and computational resources, have enabled a significant progress in seismology and have resulted in an overall picture of the properties of our planet, which has been further refined during the last couple of decades. Thus, most of the information about the Earth's internal structure comes from the detection of compressional (P) and shear (S) waves produced by earthquakes, which travel through different parts of our planet and whose velocities are sensitive to the properties of the media they traverse. These velocity distributions are then converted into density distributions via a rather non-trivial inversion problem. Indeed, this indirect technique can result in Earth models whose reliability is difficult to assess~\cite{Dziewonski20151}. Furthermore, wave velocities also depend on the composition, temperature, pressure and elastic properties of the media, which introduce uncertainties in the determination of the density profile that are difficult to estimate. Particularly difficult becomes assessing the properties of the Earth's core, which is crossed by a small fraction of the produced seismic waves that are intense enough to be detected. Therefore, the use of alternative complementary and independent methods, beyond seismic waves and gravitational measurements, could further help in our understanding of the Earth's interior.

Indeed, neutrinos are a particularly appealing tool in this regard. There are different approaches, conceptually different from previous ones, that consider neutrinos and are sensitive to properties of the Earth: the study of geoneutrinos, produced by the decay of radioactive isotopes inside the Earth, sensitive to the heat power of the Earth~\cite{Bellini:2021sow}; neutrino tomography with different neutrino beams, exploiting the effects of inelastic (absorption) or elastic forward (refraction) scattering, which depend on the nucleon and electron density profile of the Earth~\cite{Winter:2006vg}; and although not technologically feasible, studying neutrino diffraction caused by coherent neutrino scattering in the Earth~\cite{Fortes:2006, Lauter:2017uag}.

The possibility of using neutrino weak interactions to learn about the Earth's density profile through neutrino tomography was first suggested five decades ago~\cite{Placci:1973, Volkova:1974xa}. These early ideas were based on considering man-made neutrinos with energies above a few TeV, which would undergo inelastic scatterings with nucleons of the medium and get partly absorbed in their path through the Earth. The absorption probability depends on the amount and density of the matter neutrinos traverse (i.e., on the distance traveled by neutrinos within the Earth) and on their energy. Therefore, studying their attenuation for different incident angles and energies could render neutrino absorption tomography of Earth feasible, offering the possibility to constrain the Earth's density profile. Further work later extended and improved upon these ideas~\cite{Nedyalkov:1980, Nedyalkov:1981, Nedyalkov:1981b, Nedyalkov:1981yy, Nedyalkov:1983, Krastev:1983, DeRujula:1983ya, Wilson:1983an, Askarian:1985ca, Volkova:1985zc, Tsarev:1985, Borisov:1986sm, Tsarev:1986xg, Borisov:1993}, and other neutrino sources, as extraterrestrial neutrinos~\cite{Wilson:1983an, Kuo:1995, Crawford:1995, Jain:1999kp, Takahashi:2001, Reynoso:2004dt} and atmospheric neutrinos~\cite{GonzalezGarcia:2007gg, Borriello:2009ad, Takeuchi:2010, Romero:2011zzb, ICradiography, Donini:2018tsg, Salvado:2019hfn}, have also been considered since then. Forty five years after being proposed (although only a decade after the first study that considered atmospheric neutrinos to do it~\cite{GonzalezGarcia:2007gg}), the first neutrino absorption tomography was performed with actual data~\cite{Donini:2018tsg}, using the one-year sample of through-going muons produced by atmospheric muon neutrinos, collected by the IceCube neutrino observatory~\cite{TheIceCube:2016oqi}. Although still with very modest precision, the results of this first Earth tomography with neutrinos agree with those from geophysical models of the Earth's density profile, being already sensitive to the presence of the core~\cite{Donini:2018tsg, Salvado:2019hfn}. Nevertheless, one must keep in mind that these are the infant stages of neutrino tomography of Earth, which is about hundred years younger than seismic wave tomography.

On the other hand, neutrinos with energies below a few tens of GeV could experience important matter effects via coherent interactions with nucleons and electrons of the media they traverse~\cite{Wolfenstein:1977ue, Mikheyev:1985zog, Mikheev:1986wj}. Therefore, the propagation through matter could result in the modification of the neutrino oscillation pattern with respect to that corresponding to the propagation in vacuum. The details of the density profile along the neutrino path determine this pattern~\cite{Winter:2006vg}, so this is called neutrino oscillation tomography. Numerous studies have considered this method, focusing on different neutrino sources and energy ranges, with man-made beams~\cite{Ermilova:1986ph, Nicolaidis:1987fe, Ermilova:1988pw, Nicolaidis:1990jm, Ohlsson:2001ck, Ohlsson:2001fy, Jacobsson:2001zk, Winter:2005we, Minakata:2006am, Gandhi:2006gu, Wang:2010cb, Tang:2011wn, Arguelles:2012nw, Millhouse:2013qja, Asaka:2018rgk}, solar~\cite{Ioannisian:2002yj, Akhmedov:2005yt, Ioannisian:2015qwa, Ioannisian:2017dkx, Bakhti:2020tcj}, atmospheric~\cite{Agarwalla:2012uj, Rott:2015kwa, Winter:2015zwx, Bezrukov:2016gmy, Bourret:2017tkw, Naumov:2018had, Bourret:2019wme, Maderer:2021aeb, Kumar:2021faw, Kelly:2021jfs, Denton:2021rgt, Capozzi:2021hkl, Upadhyay:2021kzf, DOlivoSaez:2022vdl, Maderer:2022toi, Upadhyay:2022jfd} and supernova (SN) neutrinos~\cite{Lindner:2002wm, Akhmedov:2005yt}, but so far no neutrino detector has collected precise enough data to perform this kind of neutrino tomography. Future facilities, on the other hand, will have the required capabilities for the first determination of the Earth density profile by exploiting matter effects in atmospheric neutrino oscillations~\cite{Agarwalla:2012uj, Rott:2015kwa, Winter:2015zwx, Bezrukov:2016gmy, Bourret:2017tkw, Naumov:2018had, Bourret:2019wme, Maderer:2021aeb, Kumar:2021faw, Kelly:2021jfs, Denton:2021rgt, Capozzi:2021hkl, Upadhyay:2021kzf, DOlivoSaez:2022vdl, Maderer:2022toi, Upadhyay:2022jfd}. In this work, we show that future detectors could also probe the density profile of Earth with neutrinos from a galactic SN burst.

Earth matter effects in SN neutrinos have been studied extensively and different conclusions have been reached about their potential detection, using different SN neutrino spectra and neutrino parameters~\cite{Lagage:1987xu, Arafune:1987cj, Notzold:1987vc, Minakata:1987fj, Smirnov:1993ku, Dighe:1999bi, Lunardini:2000sw, Takahashi:2000it, Lunardini:2001pb, Takahashi:2001dc, Fogli:2001pm, Lunardini:2003eh, Dighe:2003jg, Dighe:2003vm, Dasgupta:2008my, Guo:2008mma, Scholberg:2009jr, Borriello:2012zc}. Nevertheless, the possibility to perform neutrino oscillation tomography of the entire Earth using neutrinos from a future SN explosion was only once explicitly considered two decades ago~\cite{Lindner:2002wm}. The sensitivity to the crust and upper mantle densities using this type of neutrinos was investigated in more detail a few years later~\cite{Akhmedov:2005yt}. In this work, we revisit this problem with updated neutrino oscillation parameters (such as the solar mass-squared difference, critical to determine the most sensitive energies to Earth matter effects) and also with recent SN neutrino spectra. Profiting from the expected large statistics and the low backgrounds for galactic SN neutrinos,\footnote{Notice that a SN burst lasts only for a few seconds. Consequently, backgrounds could be reduced down to a negligible level.} we compute the capabilities of future neutrino detectors to determine the density profile of the Earth. Although SN neutrinos were already detected in 1987 from a SN burst at about 50~kpc from Earth~\cite{Kamiokande-II:1987idp, Kamiokande-II:1989hkh, Bionta:1987qt, Alekseev:1987ej, Alekseev:1988gp}, neutrino detectors at the time were much smaller and only a handful of events were recorded. Future detectors will be able to detect large numbers of events with good energy resolution. The main drawbacks of using SN neutrinos are our imprecise knowledge of the arrival fluxes at Earth and the fact that only one neutrino trajectory can be used by every detector.\footnote{In strict sense, this would not be a tomography, but rather, a single-ray scan of a section of the Earth.} In this work, we shall also comment on these two issues. Furthermore, SN (and solar) neutrinos reach the Earth as mass eigenstates (in vacuum) rather than flavor eigenstates and matter effects develop at shorter distances~\cite{Akhmedov:2000cs}. Nevertheless, in this case the neutrino energy resolution of the detector determines the sensitivity to remote structures~\cite{Ioannisian:2004jk, Ioannisian:2004vv, Ioannisian:2017chl, Bakhti:2020tcj} and we shall study in detail how this affects the determination of the Earth density profile. We shall also explore the impact of energy resolution to independently constrain the core density and discuss to which extent different future neutrino detectors could actually probe the Earth's interior and whether external constraints (as the mass of the Earth) would be required to constrain the density of the inner parts of Earth. 

This article is structured as follows. In section~\ref{sec:fluxes}, we introduce the SN neutrino spectra we consider and discuss the impact of Earth matter effects on the final fluxes at detectors. In section~\ref{sec:events}, we describe the calculation of the number of events, corresponding to different detection channels, in the three future detectors we consider: DUNE~\cite{DUNE:2020lwj, DUNE:2020ypp}, Hyper-Kamiokande~\cite{Hyper-Kamiokande:2018ofw} and JUNO~\cite{JUNO:2015sjr, JUNO:2015zny, JUNO:2021vlw}. In section~\ref{sec:analysis}, we show and discuss our results for the sensitivity to the Earth density profile of the three detectors and study the impact of several factors. Finally, in section~\ref{sec:conclusions}, we summarize the results and draw our conclusions.

\section{Supernova neutrino fluxes} 
\label{sec:fluxes}

During their death, massive stars ($M\gtrsim 8\,M_\odot$) can undergo a violent explosion that is known as core-collapse SN. During this burst, $99\%$ of the energy of the star ($\sim 10^{53}$~erg) is released in the form of neutrinos of all flavors during the first seconds after the collapse. During this short time interval, three main phases can be distinguished: the neutronization burst, the accretion phase and the cooling phase. The neutronization burst, which lasts $\sim 25$~ms, consists of a huge emission of electron neutrinos due to electron capture by free protons ($e^- + p \rightarrow \nu_e + n$). The accretion phase, with a duration of $\sim 0.5$~s, is the phase in which the shock wave leads to a hot accretion mantle around the high density core of the neutron star. Finally, during the cooling phase, a neutron star is formed and neutrinos of all species are emitted, with a lower luminosity. 

In the following, we shall consider several models for the SN neutrino and antineutrino spectra at production, obtained from different numerical simulations corresponding to different progenitor masses. We then present the resulting SN neutrino fluxes, both at the Earth surface and after crossing the Earth's interior and experiencing matter effects.

\subsection{Spectra at production}

In practice, most SN neutrino spectra obtained from numerical simulations are presented using the following parameterization~\cite{Keil:2002in, Tamborra:2012ac}: 
\begin{equation}
\label{eq:emit_flux}
\phi^{0}_{\nu_\beta}(t,E_\nu) = \frac{L_{\nu_\beta}(t)}{\langle E_{\nu_\beta}\rangle (t)}\,\frac{(\alpha_{\nu_\beta}(t)+1)^{\alpha_{\nu_\beta}(t)+1}}{\langle E_{\nu_\beta} \rangle (t)\,\Gamma(\alpha_{\nu_\beta}(t)+1)}\,\left(\frac{E_\nu}{\langle E_{\nu_\beta}\rangle(t)}\right)^{\alpha_{\nu_\beta}(t)}\exp{\left(- \frac{\left(\alpha_{\nu_\beta}(t) + 1\right) E_\nu}{\langle E_{\nu_\beta}\rangle(t)}\right)} ~,
\end{equation}
which describes the differential emitted neutrino spectra for each neutrino flavor, $\nu_\beta$, at a time $t$. The luminosity of flavor $\nu_\beta$ is indicated by $L_{\nu_\beta}(t)$, $\langle E_{\nu_\beta}\rangle (t)$ is the mean neutrino energy and $\alpha_{\nu_\beta}(t)$ is the pinching parameter. Numerical simulations which compute SN neutrino spectra usually provide the results for these three parameters as a function of time, which are then introduced in Eq.~(\ref{eq:emit_flux}) to obtain the time and energy-dependent neutrino spectra. In this study, we shall use the time-integrated spectrum,
\begin{equation}\label{eq:E0flux}
    F^0_{\nu_\beta}(E_\nu) = \int^{t_\mathrm{end}}_{t_\mathrm{ini}} \phi^{0}_{\nu_\beta}(t,E_\nu)\, \mathrm{d}t~,
\end{equation}
where the initial and final emission times, $t_\mathrm{ini}$ and $t_\mathrm{end}$, are provided by the different publicly available numerical simulations (we set $t_\mathrm{ini} = 0$~s in all cases). We will also comment on the effect of time binning later on in the text. 

Core-collapse SN models depend on a wide variety of parameters (both from the physical and computational viewpoints), and the choice of parameters results in a different SN neutrino spectra. Therefore, the initial SN neutrino spectra, $F^{0}_{\nu_\beta}$, represent the most uncertain quantities in our study. Given that there is not a good agreement among the results of different simulations yet, in the following we shall take several SN models and different progenitor masses, to provide a glimpse of possible variations of our results. In this work, we consider two sets of simulations for (approximately) three different progenitor masses. We use the SN neutrino spectra as a function of time from Ref.~\cite{Warren:2019lgb}, referred to as \texttt{Warren}, parameterized according to Eq.~(\ref{eq:emit_flux}) and obtained from the repository \texttt{SNEWPY}~\cite{SNEWS:2021ezc}, which contains several other SN neutrino spectra (see also Refs.~\cite{Vartanyan:2019ssu, Burrows:2019zce, Nagakura:2020qhb, Wang:2022dva, Tsang:2022imn}). We consider three possible progenitor star masses: $9~M_\odot$, $20~M_\odot$ and $120~M_\odot$, and for turbulence strength parameter $\alpha_\Lambda =1.25$ (to guarantee successful SN explosions). We shall also use the spectra as a function of time corresponding to a $18.88~M_\odot$ progenitor star from the simulations of the Garching group~\cite{Bollig:2020phc}, referred to as \texttt{Garching19}, and also parameterized according to Eq.~(\ref{eq:emit_flux}). As our benchmark choice, whenever we quote results for a single model, we adopt the Warren $20~M_\odot$ (\texttt{Warren20}) model~\cite{Warren:2019lgb}.

\begin{figure}[t]
\begin{center}
\includegraphics[width=\columnwidth]{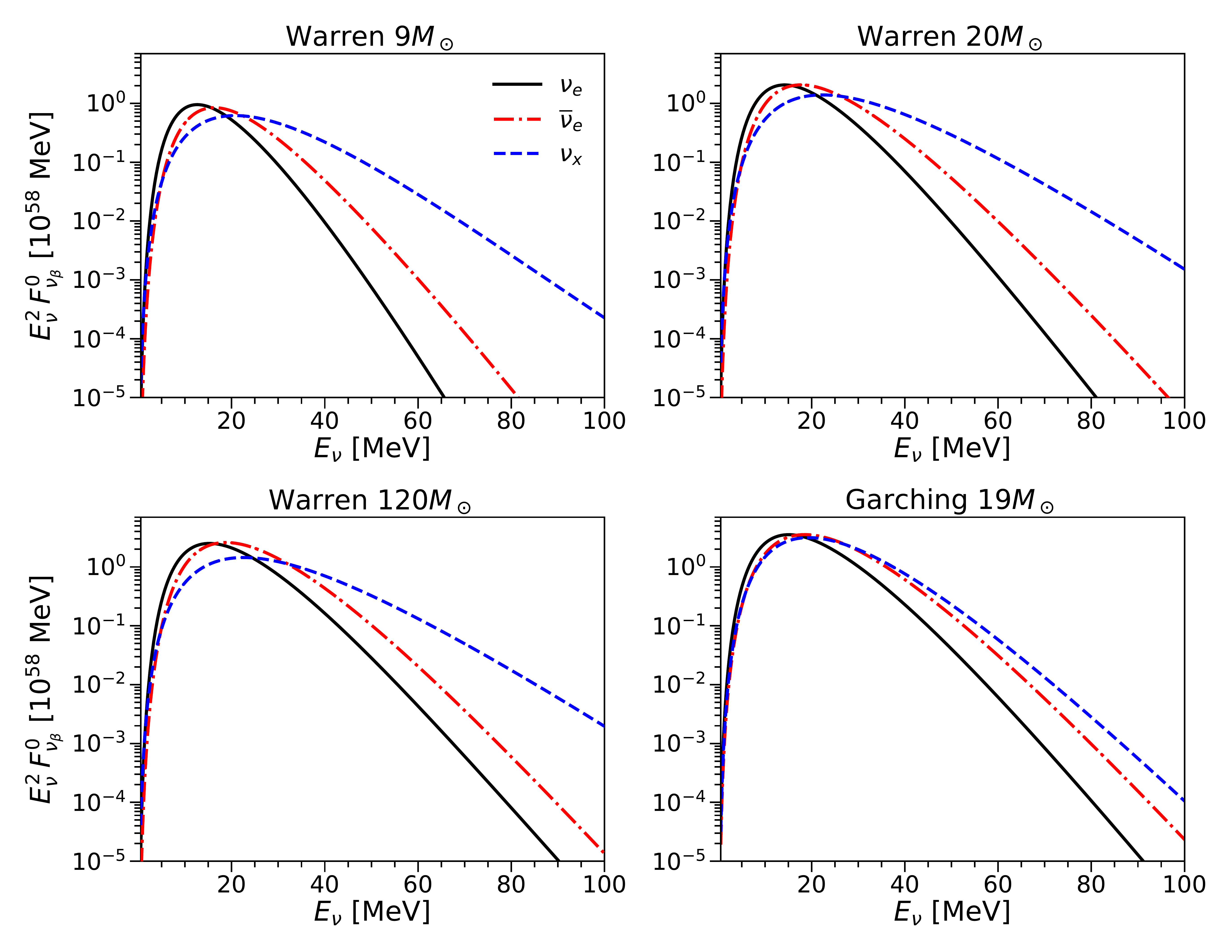} 
\caption{\textit{\textbf{Time-integrated SN neutrino spectra at production}}, as a function of energy, for the four SN simulations we consider for different progenitor masses: \texttt{Warren9}~\cite{Warren:2019lgb} (top-left panel), \texttt{Warren20}~\cite{Warren:2019lgb} (top-right panel), \texttt{Warren120}~\cite{Warren:2019lgb} (bottom-left panel) and \texttt{Garching19}~\cite{Bollig:2020phc} (bottom-right panel). In each panel, we depict the $\nu_e$ (black solid curves), $\bar{\nu}_e$ (dot-dashed red curves) and $\nu_x$ and $\bar\nu_x$ (dashed blue curves), with $x = \mu, \tau$. The $F_{\nu_x}^0$ spectra refer to the average over neutrinos and antineutrinos of $\mu$ and $\tau$ flavors.}
\label{fig:fluxes0}
\end{center}
\end{figure}

Figure~\ref{fig:fluxes0} illustrates the $\nu_e$, $\bar{\nu}_e$, $\nu_x$ (with $x=\mu, \tau$ for neutrinos and antineutrinos) time-integrated neutrino spectra provided by the different numerical simulations we consider and in Tab.~\ref{tab:param} we indicate the best-fit values of the parameters for the time-integrated spectra. Firstly, notice that the $\nu_\mu$ and $\nu_\tau$ spectra are identical, and are identified as $\nu_x$. Secondly, the $\nu_e$ initial spectrum is the highest one at the peak (as expected), since the neutronization burst is only present in this flavor, but it is the lowest one at the tail, since it is the coolest spectrum of the three. As we discuss below, one of the keys for the detectability of the Earth matter effect is the difference between the $\bar\nu_e$ (and $\nu_e$) spectrum and the $\nu_x$ one, and the smallest differences are found for \texttt{Garching19}. Lastly, notice that, while the shapes of the spectra are quite similar for all these numerical simulations, both the normalization and the peak location depend both on the mass of the progenitor star and on the simulation model.

\begin{table}[h!]
\centering
\begin{tabular}{|c||ccc|ccc|ccc|ccc|} 
\hline 
& & & & & & & & & & & & \\[-5pt]
  &  & \texttt{Warren9} & & & \texttt{Warren20} & & & \texttt{Warren120} & & & \texttt{Garching19} &  \\[5pt] 
& & & & & & & & & & & & \\[-10pt]  
  & $\nu_e$ & $\bar{\nu}_e$ & $\nu_x$ & $\nu_e$ & $\bar{\nu}_e$ & $\nu_x$ & $\nu_e$ & $\bar{\nu}_e$ & $\nu_x$ & $\nu_e$ & $\bar{\nu}_e$ & $\nu_x$ \\[5pt]
  \hline
& & & & & & & & & & & & \\[-5pt]    
 \, $L_{\nu_\beta}$ [$10^{52}$ erg] \, & \, 1.47 \, & \, 1.03 \, & \, 1.01 \, & \, 2.78 \, & \, 2.83 \, & \, 2.70 \, & \, 3.42 \, & \, 3.79 \, & \, 2.82 \, & \, 5.79 \, & \, 6.21 \, & \, 5.95 \,   \\[5pt]
 & & & & & & & & & & & & \\[-5pt]  
 $\langle E_{\nu_\beta} \rangle$ [MeV] & \, 10.6 \, & \, 13.6 \, & \, 14.4 \, & \, 11.1 \, & \, 14.8 \, & \, 16.4 \, & \, 12.0 \, & \, 15.7 \, & \, 16.6 \, & \, 11.9 \, & \, 14.7 \, & \, 14.9 \,   \\[5pt]
 & & & & & & & & & & & & \\[-5pt]   
 $\alpha_{\nu_\beta}$ & \, 2.82 \, & \, 2.90 \, & \, 1.23 \, & \, 2.12 \, & \, 2.70 \, & \, 1.50 \, & \, 2.02 \, & \, 2.84 \, & \, 1.49 \, & \, 2.10 \, & \, 2.48 \, & \, 2.14 \,  \\[5pt]
 \hline
\end{tabular}
\caption{\textbf{Best-fit parameters of the time-integrated three-parameter quasi-thermal spectra} for the four SN simulations we consider: \texttt{Warren9}~\cite{Warren:2019lgb}, \texttt{Warren20}~\cite{Warren:2019lgb}, \texttt{Warren120}~\cite{Warren:2019lgb} and \texttt{Garching19}~\cite{Bollig:2020phc}. The fit is performed to the $\log$ of Eq.~(\ref{eq:emit_flux}) so that the tail is well reproduced. In general, it is accurate (within $10\%-20\%$) for $E_\nu \gtrsim 20$~MeV.}
\label{tab:param}
\end{table}

\subsection{Fluxes at Earth}

After production in the interior of the collapsed star, SN neutrinos may undergo flavor transitions due to their coherent interactions with electrons, protons and neutrons within the SN interior, giving rise to Mikheyev-Smirnov-Wolfenstein (MSW) transitions~\cite{Wolfenstein:1977ue, Mikheyev:1985zog, Mikheev:1986wj}. Neutrinos are produced in a high-density medium, so the effective neutrino mixings are strongly suppressed and neutrinos are produced as mass eigenstates. For the current best-fit values of neutrino parameters, flavor conversions inside the star are fully adiabatic and hence, the spectra of mass eigenstates at the surface can be identified with the flavor spectra at production. For normal neutrino mass ordering (NO, $\Delta m^2_{31} > 0$), $F_{\nu_1}^0 = F_{\nu_2}^0 = F_{\nu_x}^0$, $F_{\nu_3}^0 = F_{\nu_e}^0$, $F_{\bar \nu_1}^0 = F_{\bar \nu_e}^0$ and $F_{\bar \nu_2}^0 = F_{\bar \nu_3}^0 = F_{\nu_x}^0$, while for inverted neutrino mass ordering (IO, $\Delta m^2_{31} < 0$), $F_{\nu_1}^0 = F_{\nu_3}^0 = F_{\nu_x}^0$, $F_{\nu_2}^0 = F_{\nu_e}^0$, $F_{\bar \nu_1}^0 = F_{\bar \nu_2}^0 = F_{\nu_x}^0$ and $F_{\bar \nu_3}^0 = F_{\bar \nu_e}^0$~\cite{Dighe:1999bi}. After exiting the star, neutrino mass eigenstates travel incoherently on their way to detectors on Earth, where they are detected as flavor eigenstates, whose spectra can be written as~\cite{Smirnov:1993ku, Dighe:1999bi}
\begin{eqnarray}
\label{eq:fluxD}
F^\mathrm{D}_{\nu_e} & = & p\,F^0_{\nu_e} + (1 - p) \, F^0_{\nu_x} \hspace{1cm} ; \hspace{1cm} 
F^\mathrm{D}_{\nu_x} = \frac{1 - p}{2} \, F^0_{\nu_e} + \frac{1 + p}{2} \, F^0_{\nu_x} \hspace{0.5cm} ; \nonumber \\[1ex]
F^\mathrm{D}_{\bar\nu_e} & = & \overline{p} \, F^0_{\bar\nu_e} + (1 - \overline{p}) \, F^0_{\nu_x}  \hspace{1cm} ; \hspace{1cm}
F^\mathrm{D}_{\bar\nu_x} = \frac{1 - \overline{p}}{2} \, F^0_{\bar\nu_e} + \frac{1 + \overline{p}}{2} \, F^0_{\nu_x} \hspace{0.5cm} , 
\end{eqnarray}
so that the $\nu_\beta$ fluxes at the detector are $d\Phi^\mathrm{D}_{\nu_\beta}/dE_\nu = F^\mathrm{D}_{\nu_\beta}/(4 \pi \, d_\mathrm{SN}^2)$, with $d_{\rm SN}$ the SN-Earth distance, taken throughout this work as $10$~kpc.\footnote{Notice that about 50\% of potential galactic SN is expected to occur within $\sim 10$~kpc and about 10\% within $\sim 5$~kpc~\cite{Adams:2013ana, Li:2020ujl}.} Before crossing the Earth, the survival probabilities (of the spectra at production) are given by 
\begin{eqnarray}
\label{eq:pvacuum}
p_\mathrm{vac}^\mathrm{NO} & \equiv & P_\mathrm{vac}(\nu_3 \rightarrow \nu_e) = |U_{e3}|^2 = \sin^2 \theta_{13} \hspace{1.8cm} ;  \hspace{1cm} 
p_\mathrm{vac}^\mathrm{IO} \equiv P_\mathrm{vac}(\nu_2 \rightarrow \nu_e) = |U_{e2}|^2 = \sin^2 \theta_{12} \cos^2 \theta_{13} \hspace{0.3cm} ~; \nonumber \\[1ex]
\overline{p}_\mathrm{vac}^\mathrm{NO} & \equiv & P_\mathrm{vac}(\bar{\nu}_1 \rightarrow \bar{\nu}_e) = |U_{e1}|^2 = \cos^2 \theta_{12} \, \cos^2 \theta_{13} 
\hspace{0.5cm} ; \hspace{1cm}
\overline{p}_\mathrm{vac}^\mathrm{IO} \equiv P_\mathrm{vac}(\bar{\nu}_3 \rightarrow \bar{\nu}_e) = |U_{e3}|^2 = \sin^2 \theta_{13} \hspace{0.3cm} ~.
\end{eqnarray}
From the current fits of the mixing angles~\cite{deSalas:2020pgw} (see also Refs.~\cite{Esteban:2020cvm, Capozzi:2017ipn}),\footnote{The least known parameter in the neutrino mixing matrix is the CP violating phase. Nevertheless, it does not appear in the transition probabilities (neither in vacuum nor in matter) of neutrino mass eigenstates into $\nu_e$ or $\bar\nu_e$~\cite{Ioannisian:2004vv}.} we obtain $p_\mathrm{vac}^\mathrm{NO} \simeq 0.022$,  $\overline{p}_\mathrm{vac}^\mathrm{NO} \simeq 0.67$ for NO and  $p_\mathrm{vac}^\mathrm{IO} \simeq 0.31$,  $\overline{p}_\mathrm{vac}^\mathrm{IO} \simeq 0.022$ for IO. 

Let us note that in this work we are neglecting possible non-adiabaticity effects occurring when resonances take place near the shock wave~\cite{Schirato:2002tg, Fogli:2003dw, Fogli:2004ff, Tomas:2004gr, Dasgupta:2005wn, Choubey:2006aq, Kneller:2007kg, Friedland:2020ecy}, and the presence of turbulence in the matter density~\cite{Fogli:2006xy, Friedland:2006ta, Kneller:2010sc, Lund:2013uta, Loreti:1995ae, Choubey:2007ga, Benatti:2004hn, Kneller:2013ska, Fogli:2006xy}. Inside the SN and close to the neutrinosphere, the neutrino density is so high that effects of neutrino-neutrino interactions might also be relevant and give rise to the so-called self-induced flavor conversions~\cite{Mirizzi:2015eza, Chakraborty:2016yeg, Horiuchi:2018ofe, Tamborra:2020cul}. These can be either slow~\cite{Mirizzi:2015eza, Chakraborty:2016yeg, Horiuchi:2018ofe} or fast~\cite{Tamborra:2020cul}, according to the size of the timescale for their development. Yet, there is no consensus on what the outcome of such non-linear conversions effects might be. In particular, this can range from spectral swaps to complete flavor decoherence, depending on the original energy and angular distributions of neutrinos. We refer the reader to Ref.~\cite{Choubey:2010up} for an estimate of their putative signatures in future neutrino detectors, such as DUNE. Nevertheless, a full multi-angle study of neutrino self-interactions showed that the energy-dependent modifications of the spectrum would get smeared out when considering the post-bounce time-integrated spectrum and corrections are expected to be small~\cite{Lunardini:2012ne}.

\subsection{Fluxes at the detector: Earth matter effects}
\label{sec:SNprobs}

The key to perform neutrino tomography (or neutrino scan) of Earth with SN neutrinos is to study matter effects in their propagation through the Earth, which depend on the medium density~\cite{Lagage:1987xu, Arafune:1987cj, Notzold:1987vc, Minakata:1987fj, Smirnov:1993ku, Dighe:1999bi, Lunardini:2000sw, Takahashi:2000it, Lunardini:2001pb, Takahashi:2001dc, Fogli:2001pm, Lunardini:2003eh, Dighe:2003jg, Dighe:2003vm, Dasgupta:2008my, Guo:2008mma, Scholberg:2009jr, Borriello:2012zc}. At energies of a few tens of MeV, typical of SN neutrinos, absorption in the Earth is completely negligible, but matter effects could show up in flavor transitions. At these energies, Earth matter effects are driven by the solar mass-squared difference, $\Delta m^2_{21}$. The inverse of the oscillation length (in vacuum) associated to the atmospheric sector, $\Delta m^2_{31}/(4 \pi E_\nu)$, is much larger than the effective matter potential of Earth, given by $V = \sqrt{2} \, G_F \, N_e$, with $G_F$ the Fermi constant and $N_e$ the electron number density. Thus, the $\Delta m^2_{31}-$driven oscillatory terms get averaged out and the three-neutrino calculation gets simplified to a two-neutrino problem~\cite{Kuo:1986sk}. 

An important parameter in the description of Earth matter effects of SN neutrinos is~\cite{Minakata:1987fj, Smirnov:1993ku, Ioannisian:2004jk}
\begin{equation}
\label{eq:epsilon}
\epsilon \equiv \frac{2 \, E_\nu \, V}{\Delta m^2_{21}} \simeq 0.12 \,  \left(\frac{E_\nu}{20~\mathrm{MeV}}\right) \, \left(\frac{Y_e \, \rho}{3~\mathrm{g/cm}^3}\right) \, \left(\frac{7.5 \times 10^{-5}~\mathrm{eV}^2}{\Delta m^2_{21}}\right)  ~, 
\end{equation}
where $Y_e$ is the electron fraction and $\rho$ is the matter density. It is a small parameter at energies where the peak of the SN neutrino spectra lie, where Earth matter effects only represent a small correction to the propagation in vacuum. Nevertheless, at higher energies, along the exponential tail of the SN neutrino spectra, this parameter is not small and substantial (even resonant) matter effects could take place. The oscillation length (for flavor states) in vacuum, $\ell_0$, gets modified in Earth as
\begin{equation}
\ell_\oplus = \frac{\ell_0}{\sqrt{(\cos 2\theta_{12} \mp \epsilon \, \cos^2\theta_{13})^2 + \sin^2 2\theta_{12}}} ~, 
\hspace{1cm} \mathrm{with} \hspace{0.5cm} \ell_0 \equiv \frac{4 \pi \, E_\nu}{\Delta m^2_{21}} ~,
\end{equation}
where the $- (+)$ sign corresponds to neutrinos (antineutrinos), and in general, the survival probabilities of SN neutrinos after crossing Earth are given by~\cite{Fogli:2001pm}
\begin{eqnarray}
\label{eq:pmatter}
p_\oplus^\mathrm{NO} & \equiv & P_\oplus(\nu_3 \rightarrow \nu_e) \simeq \sin^2 \theta_{13} \hspace{3.24cm} ; \hspace{1.5cm} 
p_\oplus^\mathrm{IO} \equiv P_\oplus(\nu_2 \rightarrow \nu_e) \simeq \cos^2 \theta_{13} \, P_\oplus^{2\nu}
\hspace{0.3cm} ~; \nonumber \\[1ex]
\overline{p}_\oplus^\mathrm{NO} & \equiv & P_\oplus(\bar{\nu}_1 \rightarrow \bar{\nu}_e) \simeq \cos^2 \theta_{13} \, \left( 1 - \bar{P}_\oplus^{2\nu} \right) 
\hspace{1.5cm} ; \hspace{1.5cm}
\overline{p}_\oplus^\mathrm{IO} \equiv P_\oplus(\bar{\nu}_3 \rightarrow \bar{\nu}_e) \simeq \sin^2 \theta_{13} \hspace{1cm} ~, 
\end{eqnarray}
with the two-neutrino transition probability $\nu_2 \to \nu_e$, under the approximation of constant density, given by~\cite{Smirnov:1993ku, Dighe:1999bi} 
\begin{equation}
P_\oplus^{2\nu} = \sin^2 \theta_{12} + \sin2\theta_{12}^\oplus \, \sin \left(2 \theta_{12}^\oplus - 2 \theta_{12}\right) \sin^2 \left(\pi \frac{L}{\ell_\oplus}\right) ~,
\end{equation}
where $\sin 2\theta_{12}^\oplus = (\ell_\oplus/\ell_0) \sin 2\theta_{12}$, with $\theta_{12}^\oplus$ the mixing angle of flavor states in Earth's matter, $\theta_{12}^\oplus - \theta_{12}$ the corresponding mixing angle of mass states, and $L$ the neutrino path length in the medium. The expression for $\bar{P}_\oplus^{2\nu}$ is identical, but using the appropriate oscillation length in matter for antineutrinos.

In analogy to solar neutrinos, the regeneration factor, $f_\mathrm{reg}$, can be defined as the difference between the probabilities after and before crossing Earth~\cite{Lunardini:2001pb}. It genuinely describes Earth matter effects, although in the context of SN neutrinos, unlike for solar neutrinos, it does not necessarily imply higher $\nu_e$ or $\bar{\nu}_e$ fluxes after exiting Earth, since they also depend on $F_{\nu_e}^0 - F_{\nu_x}^0$ or $F_{\bar{\nu}_e}^0 - F_{\bar{\nu}_x}^0$~\cite{Dighe:1999bi}. For $\nu_e$ and IO and for $\bar{\nu}_e$ and NO, the regeneration factor (for a constant density medium) is given by
\begin{equation}
\label{eq:freg}
f_\mathrm{reg} \equiv p_\oplus - p_\mathrm{vac} = \pm \cos^2 \theta_{13} \, \sin2\theta_{12}^\oplus \, \sin \left(2 \theta_{12}^\oplus - 2 \theta_{12}\right) \, \sin^2 \left(\pi \frac{L}{\ell_\oplus}\right) = \epsilon \, \cos^4 \theta_{13} \, \sin^22\theta_{12}^\oplus \, \sin^2 \left(\pi \frac{L}{\ell_\oplus}\right) ~,
\end{equation}
which is always positive for constant density (not necessarily so for a general density profile), since $\bar{\theta}_{12}^\oplus < \theta_{12} < \theta_{12}^\oplus$. Its amplitude has a maximum at $\epsilon \cos^2 \theta_{13} = 1$, which equals $\cos^2 \theta_{13} \cos^2 \theta_{12}$ for neutrinos and $\cos^2 \theta_{13} \sin^2\theta_{12}$ for antineutrinos~\cite{Lunardini:2000sw, Lunardini:2001pb}. Thus, the $\Delta m^2_{21}-$driven resonance in SN neutrino propagation through the Earth would take place at $E_\nu \lesssim 100$~MeV. Around the peak of the SN neutrino spectra, the solution can be obtained perturbatively, with $\epsilon$ being the perturbative parameter. In a medium of varying density and up to first order in $\epsilon$ (that is, at energies smaller than the $\Delta m^2_{21}-$driven matter resonance), $f_\mathrm{reg}$ can be approximated as~\cite{Akhmedov:2004rq, Ioannisian:2004vv}
\begin{equation}
\label{eq:fregapp}
f_\mathrm{reg} \simeq \frac{1}{2} \, \sin^2 2\theta_{12} \, \cos^4 \theta_{13} \, \int_0^L \frac{2\pi dx}{\ell_0} \, \epsilon(x) \, \sin\left( \int_x^L \frac{2 \pi dy}{\ell_\oplus(y)} \right) \leq \frac{1}{2} \, \sin^2 2\theta_{12} \, \cos^4 \theta_{13} \, \epsilon_{\rm max} ~, 
\end{equation}
where $\epsilon_{\rm max}$ is the maximum value of $\epsilon(x)$ along the neutrino path.

In short, for SN neutrinos and antineutrinos, Earth matter effects are most important at the exponentially suppressed tail of the spectrum (see Eq.~(\ref{eq:emit_flux})), where statistics would be low, whereas they are weak at the peak of the spectrum, where statistics would be high. Below, we shall discuss the energy range where the highest sensitivity to matter effects, and thus to the Earth's density profile, could be achieved by future neutrino detectors. 

In order to do so, we consider a simplified Earth's model. We take the Preliminary Reference Earth Model (PREM)~\cite{Dziewonski:1981xy} and we divide it into two major structures, the mantle and the core. We introduce a normalization factor within each region and impose the constraint on the mass of the Earth and the core to be denser than the mantle.\footnote{We could have also imposed the constraint on the Earth's moment of inertia. In that case, at least a three-layer profile would be required to have one free parameter. We choose the simpler two-layer model to better illustrate the potential of SN neutrinos to distinguish between the Earth's two major layers, the core and the mantle.} In this way, there is only one free parameter, which we choose as the normalization to the core density, $n_c$, so that 
\begin{equation}\label{eq:2layer}
    \rho_\oplus(n_c) = 
     \begin{cases}
       n_c \, \rho^\mathrm{PREM}(r)~~,~~& 0 \leq r \leq R_c ~, \\
       n_m \, \rho^\mathrm{PREM}(r)~,~~& R_c < r \leq R_\oplus ~, \\
     \end{cases}
\end{equation}  
where $R_c = 3480~\mathrm{km}$ and $R_\oplus = 6371~\mathrm{km}$ are the radius of the core and of the entire Earth, respectively, and the normalization of the mantle density, $n_m$, is fixed using the constraint on the mass of the Earth,
\begin{equation}
\label{eq:norm}
M_\oplus = \int_0^{R_c} n_c\,\rho^\mathrm{PREM}(r) \,\mathrm{d}V + \int_{R_c}^{R_\oplus}n_m\,\rho^\mathrm{PREM}(r)\,\mathrm{d}V = 5.972 \times 10^{24}~\mathrm{kg} ~.
\end{equation}

Matter effects in the propagation of SN neutrinos through the Earth depend on the neutrino trajectory and thus, on the direction of the SN relative to the detector. The path length of upward-going neutrinos inside the Earth can be given in terms of the zenith angle $\theta_z$ as $L = - 2 \, R_\oplus \, \cos \theta_z\equiv - 2 \, R_\oplus \, c_z$, with $c_z \in [-1,0)$. After neutrinos cross the Earth, the final fluxes in terms of flavor eigenstates are given by Eq.~(\ref{eq:fluxD}), with $p_\oplus$ and $\overline{p}_\oplus$ instead of $p_\mathrm{vac}$ and $\overline{p}_\mathrm{vac}$. As illustrated above, these probabilities are functions, not only of the mixing parameters, but also of the neutrino energy, of the arrival direction's zenith angle and of the Earth's density profile. 

The Earth matter effect on the SN neutrino flux is proportional to $(p_\oplus - p_\mathrm{vac}) (F_{\nu_e}^0 - F_{\nu_x}^0)$ for neutrinos and to $(\overline{p}_\oplus - \overline{p}_\mathrm{vac}) (F_{\bar{\nu}_e}^0 - F_{\bar{\nu}_x}^0)$ for antineutrinos~\cite{Dighe:1999bi}. Thus, assuming adiabatic propagation inside the star, the sensitivity to the Earth's density profile not only does depend on $f_\mathrm{reg}$, but also on the difference of neutrino spectra at production. It is important to note that, along the tail, the hardest ($\nu_e$ for NO and $\bar{\nu}_e$ for IO) fluxes at detectors correspond to the cases for which Earth matter effects are negligible. In these cases, the electron-flavor fluxes at Earth are mainly determined by the $\nu_x$ spectrum at production ($p_\oplus^\mathrm{NO} \simeq \overline{p}_\oplus^\mathrm{IO} \simeq \sin^2 \theta_{13}$).

\begin{figure}[t]
\begin{center}
\includegraphics[width=\columnwidth]{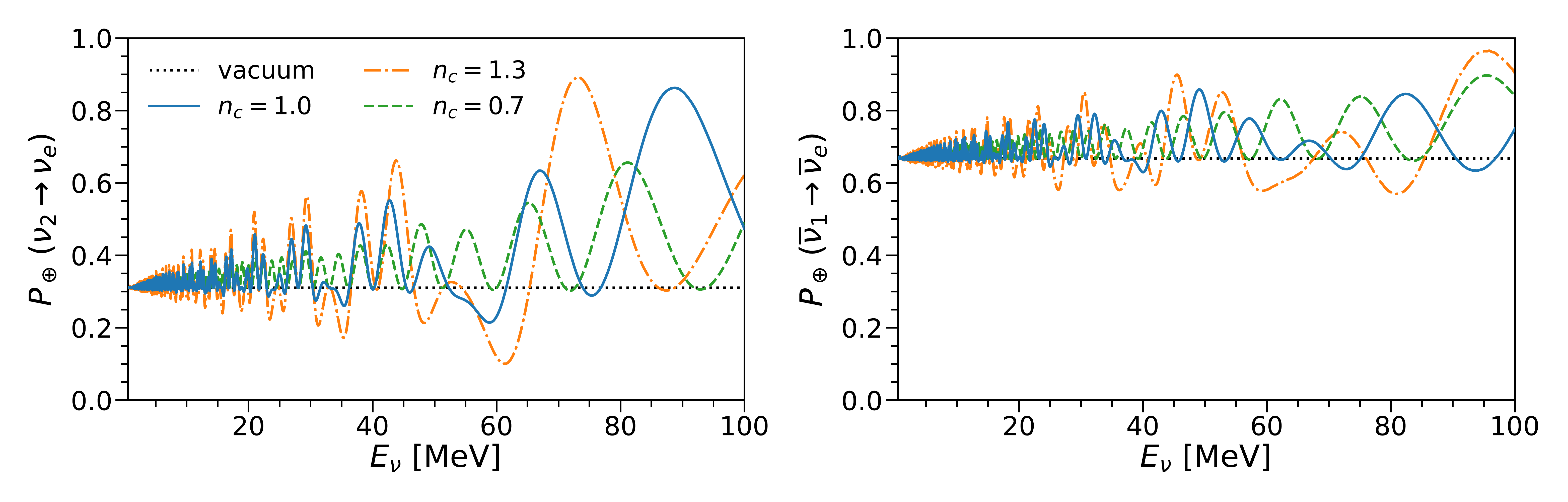}
\caption{\textit{\textbf{Transition probabilities for SN neutrinos crossing the Earth}}, for trajectories along its diameter ($\cos \theta_z = -1$), for the vacuum case (black dotted lines) and for three different profiles, parameterized by $n_c = 0.7$ (green dashed lines), $1.0$ (blue solid lines) and $1.3$ (orange dot-dashed lines). \textit{Left panel:} for neutrinos, $P_\oplus(\nu_2 \to \nu_e)$, which equals $ p_\oplus^\mathrm{IO}$, see Eq.~(\ref{eq:pmatter}), if IO and adiabatic propagation inside the SN. \textit{Right panel:} for antineutrinos, $P_\oplus(\bar{\nu}_1 \to \bar{\nu}_e)$, which equals $ \overline{p}_\oplus^\mathrm{NO}$, see Eq.~(\ref{eq:pmatter}), if NO and adiabatic propagation inside the SN.} 
\label{fig:prob}
\end{center}
\end{figure}

Figure~\ref{fig:prob} depicts the transition probabilities $P_\oplus(\nu_2 \to \nu_e)$ ($= p_\oplus^\mathrm{IO}$, if adiabatic propagation inside the SN) and $P_\oplus(\bar\nu_1 \to \bar\nu_e)$ ($= \overline{p}_\oplus^\mathrm{NO}$, if adiabatic propagation inside the SN) for neutrinos and antineutrinos crossing the entire Earth ($c_z = -1$) and for three different density profiles, parameterized by $n_c$. We also show the vacuum case for comparison. Throughout this work, the transition probabilities are obtained by means of the publicly available \texttt{nuSQuIDS} code~\cite{Arguelles:2021twb}. Transitions in matter for neutrinos in the NO case and for antineutrinos in the IO case are almost identical to the vacuum ones, making unfeasible the potential detection of Earth matter effects in these cases, and therefore they are not shown. Indeed, up to first order in $\epsilon$, the regeneration factor corresponding to these two cases is bounded by~\cite{Dighe:1999bi, Ioannisian:2004vv} 
\begin{equation}
\label{eq:diffp13}
f_\mathrm{reg} \lesssim \frac{\epsilon_{\rm max}}{2} \, \left( \frac{\Delta m^2_{21}}{\Delta m^2_{31}} \right) \,\sin^22\theta_{13} \lesssim 10^{-3} ~, 
\end{equation}
whereas the regeneration factor for neutrinos and IO, and for antineutrinos and NO, is instead bounded by
\begin{equation}
\label{eq:diffp}
f_\mathrm{reg} \lesssim \frac{\epsilon_{\rm max}}{2} \, \, \sin^2 2\theta_{12} \, \cos^4 \theta_{13} \lesssim 0.3 ~.
\end{equation}

Note that the higher the neutrino energy, the larger $\epsilon$ is, and the less accurate the linear approximation is. Indeed, matter effects are stronger along the tail of the SN spectra, where $\epsilon_{\rm max} \gtrsim 0.5$. For $E_\nu \gtrsim 50$~MeV, the regeneration factor could be larger than Eq.~(\ref{eq:diffp}), as can be seen in Fig.~\ref{fig:prob}. In what follows, we shall study the capabilities of future detectors to exploit these matter effects to perform Earth tomography, considering both neutrino mass orderings.

\section{Supernova neutrino event distributions at future neutrino detectors} 
\label{sec:events}

In this work we consider three different detector technologies, associated to three of the future neutrino detectors that will become crucial in extracting information from cosmic neutrinos with tens of MeV. Namely, we shall exploit in what follows the detection capabilities of the  liquid argon detector of the Deep Underground Neutrino Experiment (DUNE)~\cite{DUNE:2020lwj, DUNE:2020ypp}, the water-Cherenkov detector Hyper-Kamiokande (HK)~\cite{Hyper-Kamiokande:2018ofw}, and the liquid scintillator detector of the Jiangmen Underground Neutrino Observatory (JUNO)~\cite{JUNO:2015sjr, JUNO:2015zny, JUNO:2021vlw}. Whereas statistics in DUNE will be dominated by the electron neutrino flux, HK and JUNO will mainly detect electron antineutrinos. Thus, DUNE (HK and JUNO) could be mostly sensitive to Earth matter effects if IO (NO) is realized in nature, even though some sensitivity to the opposite case might be possible by considering subdominant channels. 

In all cases, the main detection channels relevant for this study occur via charged-current (CC) interactions, but the neutrino energy is reconstructed in different ways. In general, in terms of the reconstructed visible energy  $E_\mathrm{rec}$, the differential event spectrum produced by a neutrino flux $d\Phi_\nu^\mathrm{D}/dE_\nu$ is given by
\begin{equation}
\label{eq:eventrate}
\frac{dR\left(E_{\rm rec}\right)}{dE_{\rm rec}} = \, N_t \int dE_\mathrm{true} \, dE_\nu \, \varepsilon(E_{\rm true}) \, \mathcal{R}\left(E_{\rm true}, E_{\rm rec}\right) \,  \frac{d\Phi^{\rm D}_{\nu}(E_\nu)}{dE_\nu} \, \frac{d\sigma(E_\nu, E_e)}{d E_e} ~,
\end{equation}
where $N_t$ is the number of targets, which depends on the detection channel, $\varepsilon$ is the detection efficiency, and $E_\mathrm{true}$ is the true energy deposited in the detector, which depends on $E_e$, the energy of the electron/positron produced. For interactions with nuclei in DUNE, $E_\mathrm{true} = E_\nu$, and therefore the $E_\mathrm{true}$ integral is not required and the total cross section is used. For interactions with nuclei in HK and JUNO, the differential cross section of the detection channel, $d\sigma/dE_e$, is approximated as $d\sigma/dE_e = \sum_i \sigma_i \, \delta(E_\nu - E_e - Q_i - (E_x)_i)$, where $\sigma_i$ is the partial cross section of transition $i$, $Q_i$ is the corresponding mass difference between ground states of final and initial nuclei, and $(E_x)_i$ is the corresponding excitation energy of the final nucleus. The Gaussian resolution function for the visible energy, $\mathcal{R}$ in Eq.~(\ref{eq:eventrate}), is given by
\begin{equation}
\mathcal{R}\left(E_{\rm true}, E_{\rm rec}\right) = \, \frac{1}{\sqrt{2\pi} \,  \sigma_\mathrm{det}}\exp\left(-\frac{\left(E_{\rm true}-E_{\rm rec}\right)^2}{2 \, \sigma_\mathrm{det}^2}\right),
\label{eq:gauss}
\end{equation}
where $\sigma_\mathrm{det} (E_{\rm true})$ is the energy resolution of the detector in terms of the true deposited energy. Although the energy resolution is quite different among the three detectors, we consider the same energy binning (in reconstructed visible energy) for all of them, $\Delta E_\mathrm{rec} = 2$~MeV. The interval in reconstructed energy we consider goes up to 100~MeV. The minimum energy is different for each detector, although its precise value is irrelevant in this study.

\subsection{DUNE} 
\label{sec:DUNErate}

The DUNE detector~\cite{DUNE:2020lwj, DUNE:2020ypp} was originally planned as four 10-kton time projection chambers, resulting in a final mass of 40~kton of liquid argon. The total number of events is dominated by $\nu_e-$CC interactions with argon, although we also include $\bar{\nu}_e-$CC interactions with argon and $\nu-e^-$ elastic scatterings (ES) of all flavors,
\begin{align}
\nu_e\mathrm{Ar}-\mathrm{CC}: \quad \nu_e + ^{40}\mathrm{Ar} & \rightarrow e^- + \mathrm{X} ~, \nonumber \\
\bar{\nu}_e\mathrm{Ar}-\mathrm{CC}: \quad \bar{\nu}_e + ^{40}\mathrm{Ar} & \rightarrow e^+ + \mathrm{X} ~, \nonumber \\
\nu - e^- \, \mathrm{ES}: \quad \quad \, \nu + e^- & \rightarrow \nu + e^- ~. \nonumber
\end{align}

The number of argon targets is $N_t = 6.03 \times 10^{32}$ and the number of electrons is $N_e = 1.09 \times 10^{34}$. The observables of these processes are the electron/positron plus the de-excitation products of the excited final nucleus~\cite{DUNE:2020zfm}. The reconstruction of the neutrino energy requires detecting the final electron/positron, but also measuring the de-excitation photons and neutrons from the final state. The latter may produce compact and isolated low-energy features referred to as blips~\cite{ArgoNeuT:2018tvi, Castiglioni:2020tsu}. Indeed, studies of these energy depositions by low-energy photons and electrons produced after MeV neutrino-argon interactions already show, in addition to angular information, how this could help in separating $\nu_e$Ar$-$CC from $\nu-e^-$ ES events~\cite{ArgoNeuT:2018tvi, Castiglioni:2020tsu}. We assume this could be done with perfect efficiency, although this assumption has a negligible impact on our results. Thus, on one hand, we include a pure $\nu-e^-$ ES sample. On another hand, although the different energies of the de-excitation photons from each final state may also allow partial distinction between $\nu_e$Ar$-$CC and $\bar{\nu}_e$Ar$-$CC scatterings, lacking any detailed study on this regard, we conservatively consider these two channels to be indistinguishable from each other, in a single sample. As will be discussed below, given the dominance of the neutrino channel, this would not affect significantly the capabilities of DUNE to perform Earth tomography if the neutrino mass ordering is IO. If it is NO, however, isolating the subdominant $\bar{\nu}_e$ channel would be critical, although an almost perfect distinction would be required ($\gtrsim 95\%)$. Yet, even with perfect identification, given the statistics, the sensitivity to the Earth density profile would be more suppressed in the NO case. 

As the neutrino burst last only for a few seconds, different reduction cuts might be relaxed with respect to those used in searches with steady fluxes, so the detection efficiency is expected to be even better for SN neutrinos and we take $\varepsilon = 1$. For neutrino-argon interactions we assume the neutrino energy to be reconstructed in DUNE with an energy resolution of $(\sigma_\mathrm{DUNE-Ar}/E_\nu) = 0.20$~\cite{DUNE:2020ypp, Castiglioni:2020tsu}, which accounts for the electron energy resolution and for energy losses to de-excitation photons and, mainly, neutrons. For $\nu-e^-$ ES, we assume the electron kinetic energy could be reconstructed with an energy resolution of $(\sigma_\mathrm{DUNE-e}/T_e) = 0.08$. Note that, even if this energy resolution is achieved for electrons/positrons~\cite{DUNE:2020ypp},\footnote{A recent study about the reconstruction of low-energy electrons in the DUNE prototype ProtoDUNE-SP LArTPC is still far from this number, quoting $\sim 25\%$ resolution at 50~MeV~\cite{DUNE:2022meu}. Nevertheless, other studies show that calorimetric energy resolution for electron showers could reach the few per cent level~\cite{MicroBooNE:2017kvv, LArIAT:2019gdz, Castiglioni:2020tsu, MicroBooNE:2020mqg}.} energy losses from escaping particles would limit the neutrino energy resolution in neutrino-argon interactions to be worse than $\sim 15\%$ for $E_\nu > 20$~MeV~\cite{DUNE:2020ypp}. Indeed, the binding energy to free a neutron from $^{40}\mathrm{Ar}$ is 7.8~MeV and this occurs in about 15\% of $\nu_e$Ar$-$CC interactions from SN neutrinos, which corresponds to about 5\% of the total kinetic energy~\cite{Castiglioni:2020tsu}. This is one of the major factors that substantially worsen the neutrino energy resolution for $E_ \nu \gtrsim 15$~MeV. Baring all this in mind, in neutrino-argon interactions we identify the neutrino energy with the true deposited energy ($E_{\rm true} = E_\nu$) and we use the total cross section, $\sigma_\mathrm{Ar}$, in tabulated form from \texttt{SNOwGLoBES}~\cite{snowglobes}, which in turn uses the results of Refs.~\cite{GilBotella:2003sz, Kolbe:2003ys} (see also Refs.~\cite{Ormand:1994js, ICARUS:1998nzl, Bhattacharya:1998hc, Bhattacharya:2009zz, Cheoun:2011zza, Suzuki:2012ds, Karakoc:2014awa, Chauhan:2017tgf, Capozzi:2018dat}).\footnote{An event generator to obtain $\nu_e$Ar$-$CC cross sections is also available, the Model of Argon Reaction Low Energy Yields (MARLEY)~\cite{Gardiner:2020ulp, Gardiner:2021qfr}. Note that there is no direct experimental measurement of this cross section, though.} For $\nu-e^-$ ES, we consider the electron energy to be the true deposited energy ($E_\mathrm{true} = E_e$), using the differential cross section for this process~\cite{Vogel:1989iv}. 

At energies of a few to tens of MeV, the main sources of background are neutron capture processes of radiogenic and cosmogenic origin, beta decays from atmospheric muon-induced spallation products, $^8$B and \textit{hep} solar neutrinos, and atmospheric electron neutrinos CC interactions (see, e.g., Ref.~\cite{Capozzi:2018dat}). The reactor antineutrino flux is comparable to the solar one, but the smaller cross section, the higher energy threshold and the lower energies of reactor antineutrinos, render this flux irrelevant. Except for the atmospheric neutrino CC contribution, all these sources of backgrounds mainly lie at low energies ($\lesssim 20$~MeV), where matter effects are very small for Earth tomography. In any case, the short time cut of $\sim 10$~s, related to the duration of the SN burst, implies that all these backgrounds can be effectively neglected in our analyses. We choose $E_{\rm rec, min} = 5$~MeV as the minimum reconstructed neutrino energy we consider. For the Earth tomography analysis, though, its precise value is completely irrelevant. As illustrated below, even selecting only events above 20~MeV would not affect our results. Note, however, that the value of the minimum energy included in the analysis becomes important when reconstructing the SN neutrino spectra arriving at Earth.

\subsection{Hyper-Kamiokande} 
\label{sec:HKrate}

The HK detector will be the successor of the current Super-Kamiokande (SK) detector and will consist of two 220-kton water tanks~\cite{Hyper-Kamiokande:2018ofw}. In this work, because of the short duration of the signal, we take the full mass of each tank, rather than the usually assumed fiducial mass of 187~kton, considering both tanks. In water-Cherenkov detectors, the most relevant detection channels at SN  neutrino energies are inverse beta decay (IBD) ($\bar{\nu}_e$ interactions off free protons), $\nu_e-$CC and $\bar\nu_e-$CC interactions with oxygen (CC-O) and $\nu-e^-$ ES (all flavors),
\begin{align}
\mathrm{IBD}: \quad \quad \, \bar{\nu}_e + p & \rightarrow e^+ + n~,  \nonumber \\
\nu_e\mathrm{O}-\mathrm{CC}: \quad \nu_e + ^{16}\mathrm{O} & \rightarrow e^- + \mathrm{X}~,  \nonumber \\
\bar{\nu}_e\mathrm{O}-\mathrm{CC}: \quad \bar{\nu}_e + ^{16}\mathrm{O} & \rightarrow e^+ + \mathrm{X} ~, \nonumber \\
\nu - e^- \, \mathrm{ES}: \quad \, \, \, \, \, \nu + e^- & \rightarrow \nu + e^- ~. \nonumber
\end{align}

The total numbers of targets (corresponding to 440~kton) for the detection processes above are $N_t^\mathrm{IBD} = 2.94 \times 10^{34}$ protons, $N_t^\mathrm{CC-O} = N_t^\mathrm{IBD}/2 = 1.47 \times 10^{34}$ oxygen nuclei and $N_t^\mathrm{ES} = 1.47 \times 10^{35}$ electrons. As assumed for the DUNE detector, we shall also consider perfect detection  efficiencies  ($\varepsilon = 1$) for the detection of a SN neutrino burst in HK. The observed quantity in water-Cherenkov detectors is the energy of the outcoming electron/positron, assumed to be measured with a resolution as that of SK-III, $(\sigma_\mathrm{HK}/E_e) = - 0.123/(E_e/\mathrm{MeV}) + 0.0349 + 0.376/\sqrt{E_e/\mathrm{MeV}}$~\cite{Super-Kamiokande:2010tar}.\footnote{Note that energy was calibrated with electrons with energies between 4.4~MeV and 18.9~MeV~\cite{Super-Kamiokande:2010tar}.} For IBD, at energies well below the targets mass, the true deposited energy ($E_{\rm true} = E_e$) is approximately given in terms of the incoming antineutrino energy, $E_e \simeq E_\nu - (m_n - m_p)$. This might be a reasonable approximation for the bulk of SN neutrinos, since most of the events are expected to have energies $E_\nu \lesssim 40$~MeV. In the case of Earth tomography, however, most of the sensitivity lies at $E_\nu \gtrsim 40$~MeV and neutrino energy reconstruction becomes crucial to identify matter effects. Therefore, one-to-one relations between the (anti)neutrino energy and the electron (positron) energy neglect the important smearing of the latter, which is a rough approximation for IBD and for interactions with nuclei~\cite{Kolbe:1992xu}. In the case of $\nu-e^-$ ES there is not such a relation, so the differential cross section must be used.

For IBD, the energy distribution of the final positron has a width of ${\cal O}(2 E_\nu/m_p)$. This energy spread represents the limit on how well the neutrino energy could be reconstructed. Given that  for energies of a few tens of MeV this width could be larger than the experimental energy resolution, we shall use the full IBD differential cross section~\cite{Strumia:2003zx, Ricciardi:2022pru}, instead of the total cross section, so that nucleon recoils are accounted for. For interactions with oxygen, relevant at the highest energies of this work, we consider the fit to the four-group case (42 final states combined into four transitions with different excitation energies) given in Ref.~\cite{Nakazato:2018xkv} (see also Refs.~\cite{Haxton:1987kc, Fukugita:1988hw, Kuramoto:1989tk, Kolbe:1992xu, Nussinov:2000qc, Kolbe:2002gk, Botrugno:2005kn, SajjadAthar:2005ke, Lazauskas:2007bs, Chauhan:2017tgf, Suzuki:2018aey}) and obtained from shell model calculations~\cite{Suzuki:2008zzi} with a modified version of the SFO Hamiltonian~\cite{Suzuki:2003fw}. At neutrino energies where Earth matter effects are more important, $E_\nu \sim (40 - 100)$~MeV, the cross section is dominated by transitions to excited states. For each of the four transitions, the energy of the outcoming electron/positron is given by $E_e = E_\nu - Q - E_x$, with $Q_{^{16}\mathrm{O}^{16}\mathrm{F}} = 14.90$~MeV (for $\nu_e$O$-$CC interactions) and $Q_{^{16}\mathrm{O}^{16}\mathrm{N}} = 10.93$~MeV (for $\bar{\nu}_e$O$-$CC interactions). Although interactions with oxygen nuclei represent a small fraction of the total number of events, they are the most important ones to study matter effects if the neutrino mass ordering is IO, since the sensitivity to the matter potential relies on $\nu_e$O$-$CC interactions. Therefore, accounting for the electron/positron energy distribution in this case becomes critical.

We assume the two detector tanks will be doped with Gadolinium~\cite{Beacom:2003nk} with a final concentration of 0.1\%, along with the use of better photosensors, which would significantly improve the efficiency for neutron tagging~\cite{Hyper-Kamiokande:2018ofw}. In this way, we assume an identification efficiency of IBD events of 90\%, corresponding to the neutron capture efficiency with the assumed Gd concentration. In our analyses, in addition to these identified IBD events, we also include the sample composed of unidentified (10\% of the total) IBD events plus all events from $\nu_e$ and $\bar{\nu}_e$ CC interactions with oxygen and from $\nu-e^-$ ES. Matter effects will be significant in the case of $\bar{\nu}_e$ ($\nu_e$) and NO (IO). Thus, most of the power of the analyses comes from the IBD sample in the NO case, whereas the CC-O sample is the most important one in the IO case. In both  NO and IO cases the $\nu-e^-$ ES is a subdominant channel at energies where matter effects are important. Furthermore, it has a pronounced forward peak which can help distinguishing it from the remaining detection channels. In addition, note that, at tens of MeV, partial separation between neutrino and antineutrino interactions, both off free protons and off oxygen, might be possible by making use of their angular distributions. The former have a backward peak, whereas the latter have a forward peak, which become more pronounced as energy grows~\cite{Kuramoto:1989tk}. We do not account for this possibility in this work, though.

In our analyses, we use a minimum reconstructed energy of $E_{\rm rec, min} = 3$~MeV~\cite{Hyper-Kamiokande:2018ofw}. As mentioned above, however, at low energies matter effects are small, so low-energy bins could be removed from the analysis without affecting the final results. Assuming efficient neutron tagging, spallations products with accompanying neutrons, accidental coincidences with spallation products and reactor antineutrinos are the main sources of background for $E_e \lesssim 12$~MeV.\footnote{The background from neutral-current (NC) interactions of atmospheric neutrinos is significant up to slightly higher energies, although this background can be efficiently removed by accounting for different intervals of the opening angle of the Cherenkov light cone~\cite{Bays:2011si, Bays:2012wty, Super-Kamiokande:2021jaq}.} At energies $E_e \gtrsim 12$~MeV, Michel electrons/positrons from invisible muons produced by atmospheric muon neutrinos/antineutrinos and atmospheric electron neutrino/antineutrino CC events are the most important sources of background. Nevertheless, the short duration of the signal allows the reduction of all these background rates to negligible levels, so we assume a background-free SN neutrino detection at HK.

\subsection{JUNO} 
\label{sec:JUNO}

The JUNO detector is a planned liquid scintillator detector with a total mass of 20~kton~\cite{JUNO:2015sjr, JUNO:2015zny, JUNO:2021vlw}. As for water-Cherenkov detectors, the most important detection channel at these energies is IBD, but there are also subdominant CC interactions with carbon nuclei (CC-C) and $\nu-e^-$ ES (all flavors),
\begin{align}
\mathrm{IBD}: \quad \quad \,  \bar{\nu}_e + p & \rightarrow e^+ + n ~,  \nonumber \\
\nu_e\mathrm{C}-\mathrm{CC}: \quad \nu_e + ^{12}\mathrm{C} & \rightarrow e^- + \mathrm{X} ~,  \nonumber \\
\bar{\nu}_e\mathrm{C}-\mathrm{CC}: \quad \bar{\nu}_e + ^{12}\mathrm{C} & \rightarrow e^+ + \mathrm{X} ~, \nonumber \\
\nu - e^- \, \mathrm{ES}: \quad \, \, \, \, \, \nu + e^- & \rightarrow \nu + e^- ~. \nonumber
\end{align}

Although in analyses of steady fluxes, volume cuts would reduce the fiducial mass of the detector, in the detection of neutrinos from a SN burst, the full mass could be considered. Assuming a linear alkylbenzene based liquid scintillator ($\mathrm{C}_6 \mathrm{H}_5 \mathrm{C}_{12} \mathrm{H}_{25}$), the number of free protons is $N_t^\mathrm{IBD} = 1.47 \times 10^{33}$~\cite{JUNO:2015zny}, the number of carbon nuclei is $N_t^\mathrm{CC-C} = 8.80 \times 10^{32}$ and the number of electrons is $N_t^\mathrm{ES} = 6.75 \times 10^{33}$. As for the other detectors, we also assume perfect detection efficiency, $\varepsilon = 1$, for each of the four detection channels above. The way of identifying the neutrino-induced events in JUNO is different from water-Cherenkov detectors. Rather than detecting the Cherenkov light from the final electron/positron, in the JUNO detector, photons from both ionization energy losses and electron-positron annihilations could be tagged. Hence, in IBD or $\bar{\nu}_e\mathrm{C}-$CC interactions the total deposited energy ($E_{\rm true} = E_\mathrm{dep}$) is the sum of both photon energies ($E_\mathrm{dep} = E_e + m_e$), whereas for $\nu_e\mathrm{C}-$CC interactions, only the kinetic energy of the electron is reconstructed ($E_\mathrm{dep} = E_e - m_e$). For $\nu-e^-$ ES, the reconstructed quantity is also the kinetic energy of the electron ($E_\mathrm{dep} = E_e - m_e$). The deposited energy resolution is taken to be $(\sigma_\mathrm{JUNO}/E_\mathrm{dep})^2 = 0.0261^2/(E_\mathrm{dep}/\mathrm{MeV}) + 0.0082^2 + (0.0123/(E_\mathrm{dep}/\mathrm{MeV}))^2$~\cite{JUNO:2020xtj, JUNO:2021vlw}.\footnote{Note that energy calibration was performed with energies up to 12~MeV~\cite{JUNO:2020xtj}.} 

In comparison to water-Cherenkov detectors as HK or SK, JUNO will have superb energy resolution, which a priori, would make it specially suitable for Earth tomography studies with SN neutrinos. Nevertheless, although this allows for partial compensation of the smaller size of the detector, the intrinsic energy distribution of the final electron/positron sets the reach limit of the detector. Therefore, given the dominance of the IBD channel, using the full IBD differential cross section also becomes of crucial importance in the case of JUNO, in order to properly account for the energy distribution of the outcoming positron. 

Interactions with carbon nuclei and $\nu-e^-$ ES comprise only a small fraction of the total number of events, but $\nu_e-$CC interactions with carbon nuclei are the most sensitive ones to matter effects if the neutrino mass ordering is IO. Thus, it is also important to account for the final electron/positron energy distribution in those cases. For the cross section for interactions with carbon, we use the results of Ref.~\cite{Yoshida:2008zb} (see also Refs.~\cite{OConnell:1972edu, Fukugita:1988hg, Kolbe:1999au, Hayes:1999ew, Volpe:2000zn, Botrugno:2005kn, SajjadAthar:2005ke, Paar:2007fi, Cheoun:2010zzb, Samana:2010up, Chauhan:2017tgf, Jachowicz:2021ieb}), which are obtained by shell model calculations with the SFO Hamiltonian~\cite{Suzuki:2003fw, Suzuki:2006qd}. At the energies of interest for this work, $E_\nu \lesssim 100$~MeV, more than $90\%$ of the strength in both neutrino and antineutrino interactions comes from two channels: $^{12}\mathrm{C} (\nu_e, e^-)^{12}\mathrm{N}$, $^{12}\mathrm{C} (\nu_e, e^- \, p)^{11}\mathrm{C}$, and $^{12}\mathrm{C} (\bar{\nu}_e, e^+)^{12}\mathrm{B}$, $^{12}\mathrm{C} (\bar{\nu}_e, e^+ \, n)^{11}\mathrm{B}$. Whereas at energies around the peak of the SN neutrino event spectrum, the transition to the ground state of $^{12}\mathrm{N}$ (for $\nu_e$C$-$CC) or $^{12}\mathrm{B}$ (for $\bar{\nu}_e$C$-$CC) is the dominant one, at $E_\nu \gtrsim 50$~MeV, transitions to excited states are the most important ones. In particular, spin-dipole transitions to the $2^-$ and $1^-$ states are the ones with largest strengths, with $0^-$ states contributing to the cross section at the few percent level~\cite{Gaarde:1984tpk}. In this work, we include the transition to the ground state and we combine the contributions to excited states into transitions to a single $1^-$ and a single $2^-$ state, with equal weights on the total cross section~\cite{Gaarde:1984tpk}, and with excitation energies of $\sim 4$~MeV ($2^-$ transitions) and $\sim 7.5$~MeV ($1^-$ transitions)~\cite{Gaarde:1984tpk, Ichihara:1994swx}. More precisely, for the excitation energies we take: $(E_x)_{^{12}\mathrm{N}}^{2^-} = 4.14$~MeV, $(E_x)_{^{12}\mathrm{N}}^{1^-} = 7.40$~MeV, and $(E_x)_{^{12}\mathrm{B}}^{2^-} = 4.46$~MeV, $(E_x)_{^{12}\mathrm{B}}^{1^-} = 7.70$~MeV~\cite{Ajzenberg-Selove:1990fsm, Anderson:1996gva, Schopper:2012gzf}. The energy of the final electron/positron is given by $E_e = E_\nu - Q - E_x$, where the corresponding mass differences between ground states of the final and initial nuclei are $Q_{^{12}\mathrm{C}^{12}\mathrm{N}} = 16.83$~MeV (for $\nu_e$C$-$CC interactions) and $Q_{^{12}\mathrm{C}^{12}\mathrm{B}} = 13.88$~MeV (for $\bar{\nu}_e$C$-$CC interactions).

In the JUNO detector, tagging the prompt annihilation of the positron and the delayed gamma emission after neutron capture by protons will allow identifying the IBD channel with great efficiency. Thus, we consider the IBD sample assuming an identification efficiency of 95\%. On another hand, the delayed-prompt coincident signals from the beta decays of $^{12}\mathrm{N}$ and $^{12}\mathrm{B}$ might also allow distinguishing $\nu_e\mathrm{C}-$CC from $\bar{\nu}_e\mathrm{C}-$CC interactions. Nonetheless, we conservatively combine them into a single sample, together with the 5\% of unidentified IBD events. 

Limited only by the intrinsic radioactive elements of the liquid itself, the energy threshold in JUNO could be as low as 0.1~MeV~\cite{Fang:2019lej}. As already mentioned, the precise value chosen for the minimum reconstructed energy is not important for the present study, therefore we set it equal to that of HK, $E_{\rm rec, min} = 3$~MeV. Finally, the most important backgrounds for IBD are reactor antineutrinos, geoneutrinos and atmospheric neutrino events. Other backgrounds as natural radioactivities and cosmogenic isotopes are even less important. Nevertheless, as for the other detectors, the short duration of the SN burst, $\sim 10$~s, makes all these backgrounds completely negligible.

\subsection{Event distributions at DUNE, HK and JUNO} 
\label{sec:Warren20dist}

Before showing some examples of expected event distributions, other channels beyond the main detection ones, which could be important for neutrino spectra reconstruction, deserve some general comments within the context of our analysis. In the case of neutrino-proton elastic scattering~\cite{Beacom:2002hs}, the reconstructed spectrum would lie at lower energies than for the main detection channels, mostly below detection threshold. Thus, the number of events would be much smaller and they would lie at energies for which matter effects are significantly less important. This channel would have a negligible impact when performing Earth tomography with SN neutrinos, so we have not included it in our analyses. On another hand, NC process are not sensitive to neutrino oscillations, being only sensitive to the total flux. They could, however, partly wash out matter effects on the dominant CC channels if they cannot be identified. Even if this were the case, they would only affect well defined low-energy bins, which contribute negligibly to the determination of the Earth density profile. In particular, in DUNE, a 9.8~MeV Ar$^*$ decay line would likely be the most important one~\cite{DUNE:2020zfm}, although this has not been investigated in detail. In HK, the expected de-excitation photons from the NC subdominant channel would have energies of $\sim 5-6$~MeV~\cite{Langanke:1995he}, which after Compton scattering or electron-positron pair production, would typically result in final visible energies below 5~MeV. In JUNO, NC interactions would result in photons with energies $\lesssim 15$~MeV~\cite{Lu:2016ipr}.

\begin{figure}[t]
\begin{center}
\includegraphics[width=\columnwidth]{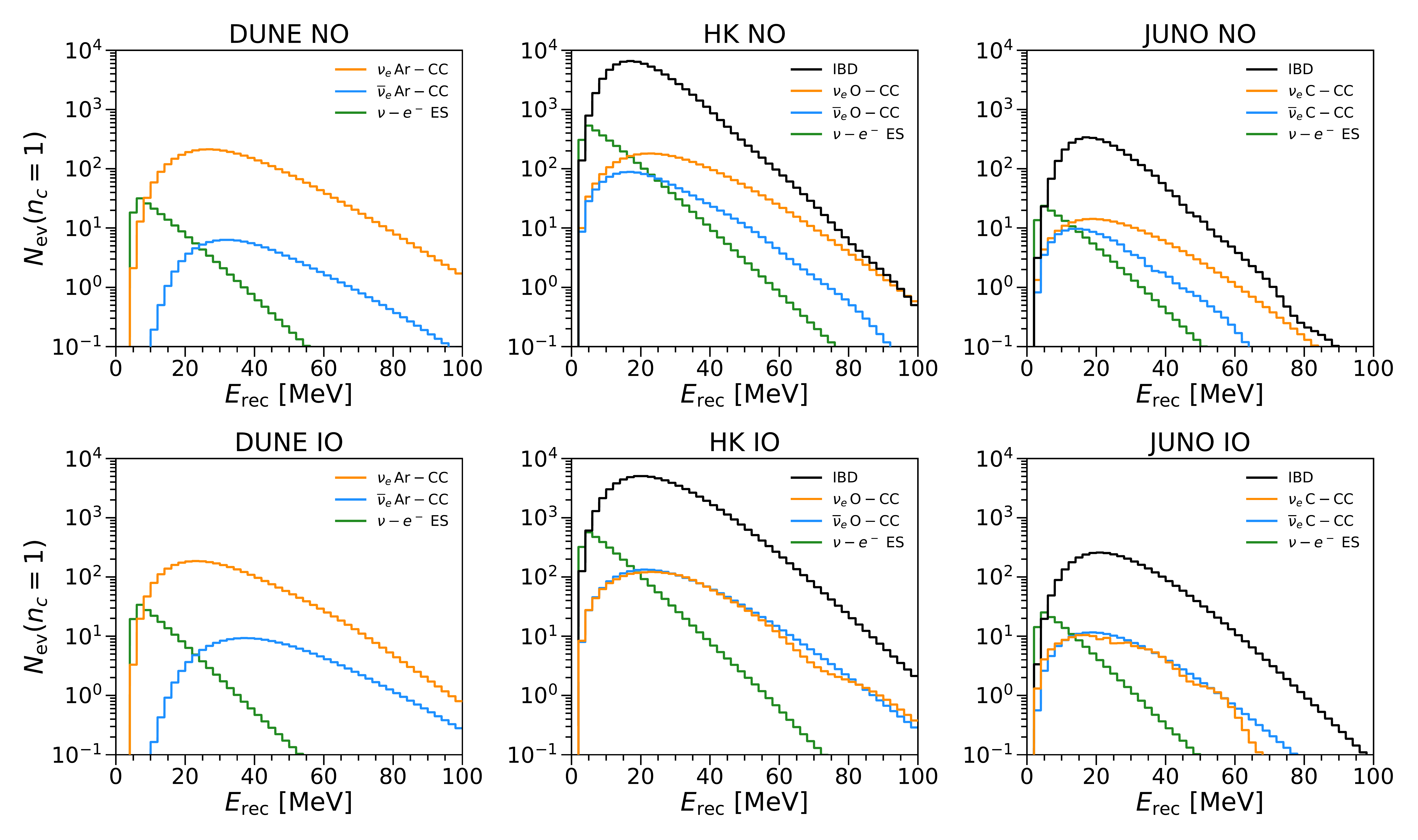}
\caption{\textit{\textbf{Event distributions}}, corresponding to the \texttt{Warren20} SN neutrino spectra~\cite{Warren:2019lgb}, as a function of the reconstructed energy, at DUNE (left panels), HK (middle panels) and JUNO (right panels), for both neutrino mass orderings: NO (top panels) and IO (bottom panels). Results are shown for all detection channels considered in this work: IBD (black lines), $\nu_e$ (orange lines) and $\bar{\nu}_e$ (blue lines) CC interactions with nuclei, and $\nu-e^-$ ES (green lines). We assume adiabatic propagation inside the SN, the SN-Earth distance to be $10$~kpc and the SN burst to occur on the opposite side of the detector (i.e., $c_z = - 1$). The Earth's density distribution is given by the PREM profile (i.e., $n_c = 1$).}
\label{fig:totalrate}
\end{center}
\end{figure}

Therefore, including all these channels in our analyses would not modify our conclusions, and in the following we shall focus on the ones indicated in the previous subsections.  Figure~\ref{fig:totalrate} shows the expected event energy distributions, for all detection channels indicated above, at the three detectors we study: DUNE, HK and JUNO. We take the \texttt{Warren20} model for the SN neutrino spectra~\cite{Warren:2019lgb} and $c_z = -1$ as an example and, as we do throughout this work, we set the SN-Earth distance to 10~kpc. We compute every event distribution using the PREM density profile (i.e., $n_c = 1$). Assuming adiabatic propagation inside the SN, substantial matter effects would occur in the propagation through the Earth of neutrinos or antineutrinos depending on the neutrino mass ordering. Thus, each detector will be more sensitive to the density profile for one ordering or the other depending on its main detection channel.

Matter effects in neutrino transitions within the Earth could be important (even resonant) for $\nu_e$ if IO, whereas they could be significant for $\bar\nu_e$ if NO. This can be seen from the figure, more pronouncedly in the case of JUNO, due to its better energy resolution. Matter effects instead would be negligible for $\nu_e$ in the case of NO and for $\bar{\nu}_e$ in the case of IO~\cite{Dighe:1999bi}, making these cases hopeless to perform Earth tomography with SN neutrinos. Nevertheless, given that the highest (electron-flavor) fluxes would correspond to the cases for which matter effects are negligible, the distinction among detection channels becomes very important to avoid much dilution of the matter effect in data. As mentioned above, in addition to the main detection channel for each detector, we include $\bar{\nu}_e\mathrm{Ar}-$CC and $\nu-e^-$ES in DUNE, $\nu_e\mathrm{O}-$CC, $\bar{\nu}_e\mathrm{O}-$CC and $\nu-e^-$ES in HK, and $\nu_e\mathrm{C}-$CC, $\bar{\nu}_e\mathrm{C}-$CC and $\nu-e^-$ES in JUNO. The inclusion of these channels does not alter our conclusions for both orderings in DUNE, and for NO in HK and JUNO, but it is required to study the sensitivity of HK and JUNO to matter effects in the IO case.

\begin{figure}[t]
\begin{center}
\includegraphics[width=\columnwidth]{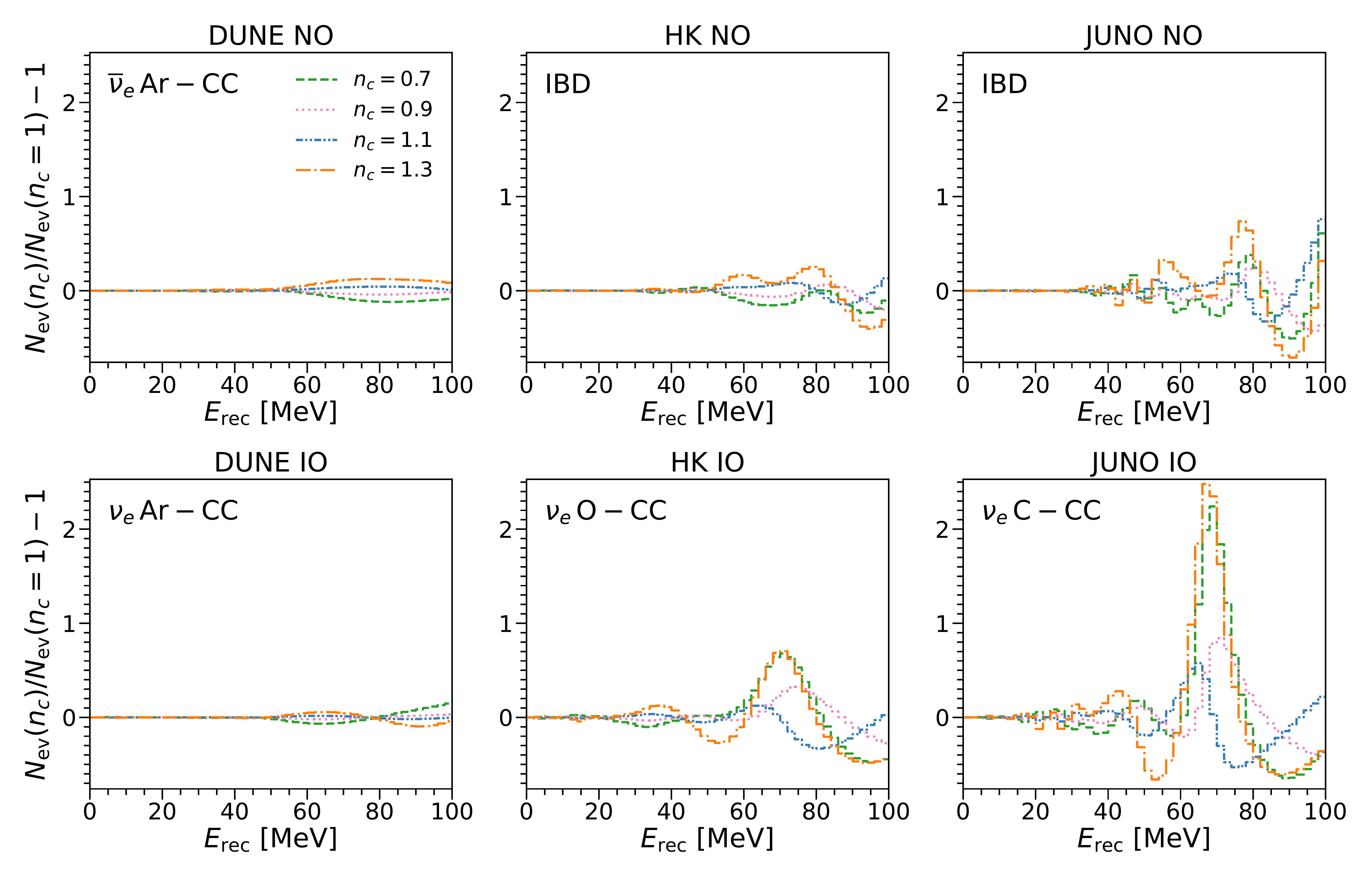}
\caption{\textit{\textbf{Relative deviations in the number of events with respect to the PREM expectation}}, corresponding to the \texttt{Warren20} SN neutrino spectra~\cite{Warren:2019lgb}, as a function of the reconstructed energy, at DUNE (left panels), HK (middle panels) and JUNO (right panels), for both neutrino mass orderings: NO (top panels) and IO (bottom panels). Results are depicted for different density profiles parameterized by the core density: $n_c = 0.7$ (green dashed curves), $n_c = 0.9$ (magenta dotted curves), $n_c = 1.1$ (blue dot-dot-dashed curves) and $n_c= 1.3$ (orange dot-dashed curves). 
Results are only shown for events produced by the detection channel most sensitive to matter effects: $\bar{\nu}_e\mathrm{Ar}-$CC ($\nu_e\mathrm{Ar}-$CC) for NO (IO) at DUNE, IBD ($\nu_e\mathrm{O}-$CC) for NO (IO) at HK, and IBD ($\nu_e\mathrm{C}-$CC) for NO (IO) at JUNO. We assume adiabatic propagation inside the SN and the SN burst to occur on the opposite side of the detector (i.e., $c_z = - 1$).}
\label{fig:reldevrate}
\end{center}
\end{figure}

Clearly, the detector with the highest statistics is HK, with $\sim 20$ times more free protons than JUNO and $\sim 50$ times more targets than DUNE. The scaling of the total number of events between HK and JUNO, given the same main detection channel (IBD), is  (approximately) proportional to the mass of the detectors. The event statistics in DUNE would be slightly lower than in JUNO, even being twice as massive. This difference can be explained from the fact that DUNE will have $\sim 2.4$ times fewer targets than JUNO. In turn, this is partly compensated by the cross section of $\nu_e\mathrm{Ar}-$CC interactions, which is slightly larger than that of IBD around the peak of the spectrum, and by the differences in neutrino and antineutrino fluxes. Along the tail of the distribution, however, where matter effects start playing a significant role, the relative contributions do not correspond to those of the totals. The growing $\nu_e\mathrm{Ar}-$CC cross section with energy and the fact that the $\nu_x$ contribution for neutrinos and IO is larger than for antineutrinos and NO, make the ratio of the number of events in DUNE (for IO) to that in HK or JUNO (for NO) to grow with energy. For instance, for the \texttt{Warren20} spectra, although the total number of events in JUNO would be larger than in DUNE, for energies $\gtrsim 40$~MeV the number of events in DUNE (IO) is expected to be more than three times larger than in JUNO (NO). Note, however, that for these spectra and the assumed SN-Earth distance, above $\sim 80$~MeV at DUNE (IO) and HK (NO) and above $\sim 60$~MeV at JUNO (NO), the number of events per bin would be just a few or smaller. For the opposite mass ordering, matter effects take place in subdominant channels, so the energies above which the number of events (produced by these channels) is just a few or smaller are lower ($\sim 40$~MeV for DUNE (NO) and JUNO (IO) and $\sim 70$~MeV for HK (IO)). Finally, notice that events produced by $\nu-e^-$ ES lie at low energies, where matter effects are small, and therefore they have a very small impact on our results. 

In Fig.~\ref{fig:reldevrate} we show the relative variations of the event distributions with respect to the PREM case ($n_c = 1$), for different density profiles (parameterized by $n_c$) and for the three detectors. Notice that larger departures of $n_c$ from its PREM value result in larger relative differences. This is a sign of the importance of matter effects, which are quite small below $\sim 40$~MeV, but grow with energy. Nonetheless, even if matter effects are less important at lower energies, the statistics would be much higher, so one could wonder whether any feature could be identified around the peak of the spectrum, $E_\nu \sim (10 - 20)$~MeV. In principle, the modulation of the neutrino flux at those energies would mark the presence of matter effects~\cite{Dighe:2003jg}. Notwithstanding, the fast induced oscillations would get greatly smeared out by the neutrino energy resolution at detection. On the other hand, at the highest energies, the number of events per bin would be very small. For instance, for some of the cases in Fig.~\ref{fig:reldevrate}, the relative growth at $\lesssim 100$~MeV occurs in bins with almost no events, with very little statistical weight. Therefore, most of the sensitivity to the density profile would lie in the energy interval $E_\nu \sim (40 - 80)$~MeV. 

The detector with the largest relative differences is JUNO, followed by HK and DUNE. This can be understood from the energy resolution of each detector and from the intrinsic width of the energy distribution of the outgoing electron/positron. The better the energy resolution, the better matter effects could be resolved and the smaller the potential washed out effect would be. However, even if the visible energy resolution is much better than the intrinsic positron energy spread in IBD, which is $\sim {\cal O}\left(2 \, E_\nu/m_p\right)$, the neutrino energy could not be reconstructed with better precision than the latter. Therefore, the sensitivity to matter effects occurring beyond the attenuation length related to the width of the positron energy distribution is greatly reduced (see section~\ref{sec:energyresolution}). On another hand, the relative deviations are not symmetrical with respect to the PREM profile. The explanation lies on the interplay between the differences in densities and the neutrino energy resolution. The larger the core density, the larger the average density along the traversed path and thus, the lower the energies for which matter effects start being important. However, the sensitivity to remote structures of a detector is driven by its capability to reconstruct the neutrino energy~\cite{Ioannisian:2004jk, Ioannisian:2004vv, Ioannisian:2017chl, Bakhti:2020tcj}. The worse this is, the less precise the core could be resolved, so matter effects would be dominated by regions closer to the detector. Therefore, the average density that could be probed would actually be effectively lower and matter effects start being important for higher energies, where statistics is lower, as detailed in the next section.

\section{Sensitivity to the Earth density profile} 
\label{sec:analysis}

In order to study the sensitivity of DUNE, HK and JUNO to matter effects in the propagation of SN neutrinos through the Earth and to establish their potential capabilities to provide information about the Earth's density profile, we assume that the incoming SN neutrino flux could be extracted, either by another detector on a different location or by using data around the peak of the SN spectrum, where most events would lie and where matter effects are very suppressed. For instance, a detector like HK will be able to distinguish among different SN neutrino models with high confidence, even for SN explosions occurring farther away than what is assumed here~\cite{Hyper-Kamiokande:2021frf}. Even if the tail of the spectra, where Earth matter effects become important, is not measured at a different location, it could still be possible to observe these effects at a single detector. Characteristic frequencies in the inverse-energy spectrum, independent of the initial SN neutrino spectra, could be identified via Fourier analysis~\cite{Dighe:2003jg, Dighe:2003vm, Akhmedov:2005yt, Borriello:2012zc}. Also note that the high-energy tails of SN neutrino spectra are expected to be described with good accuracy in terms of the simple three-parameter quasi-thermal fit we consider in this work, Eq.~(\ref{eq:emit_flux})~\cite{Tamborra:2012ac}.

In what follows, we use the time-integrated SN energy spectra and compute the event distributions for the different detection channels in each detector, combining them into two samples, as described above. We perform a binned extended maximum likelihood analysis and use the likelihood ratio as our test statistics. Assuming no background and that the two data samples could be distinguished from each other, we can write the total log-likelihood-ratio as,
\begin{equation}
\label{eq:chi2}
\Delta \chi^2 (n_c \,;c_z) = 2 \, \sum_{i,s} \left[ N_{i, s}(n_c \,; c_z) - N_{i, s}(n_c = 1 \,; c_z) + N_{i, s}(n_c = 1 \,; c_z) \, \ln\left(\frac{N_{i, s}(n_c = 1 \,; c_z)}{N_{i, s}(n_c \,; c_z)}\right)\right] ~,
\end{equation}
where $N_{i, s}(n_c \,; c_z)$ is the expected number of events in the energy bin $i$ for sample $s$, assuming a density profile characterized by $n_c$ and with the incoming neutrino direction determined by $c_z$. The expected number of events in sample $s$ in the energy bin $i$ for the PREM profile is given by $N_{i,s}(n_c = 1 \,; c_z)$ (Asimov data set). We fix the neutrino mixing parameters and the mass-squared differences to their current global best-fit values~\cite{deSalas:2020pgw}. In order to make the dependence on these parameters explicit, we shall discuss it below, rather than including a penalty term in the likelihood above and profiling over those parameters. In this regard, most of our results are slightly optimistic given current uncertainties in the neutrino parameters. Finally, we assume that the log-likelihood-ratio, Eq.~(\ref{eq:chi2}), follows a $\chi^2$ distribution with one degree of freedom.

\subsection{Dependence on the SN neutrino spectra}

\begin{figure}[t]
\begin{center}
\includegraphics[width=\columnwidth]{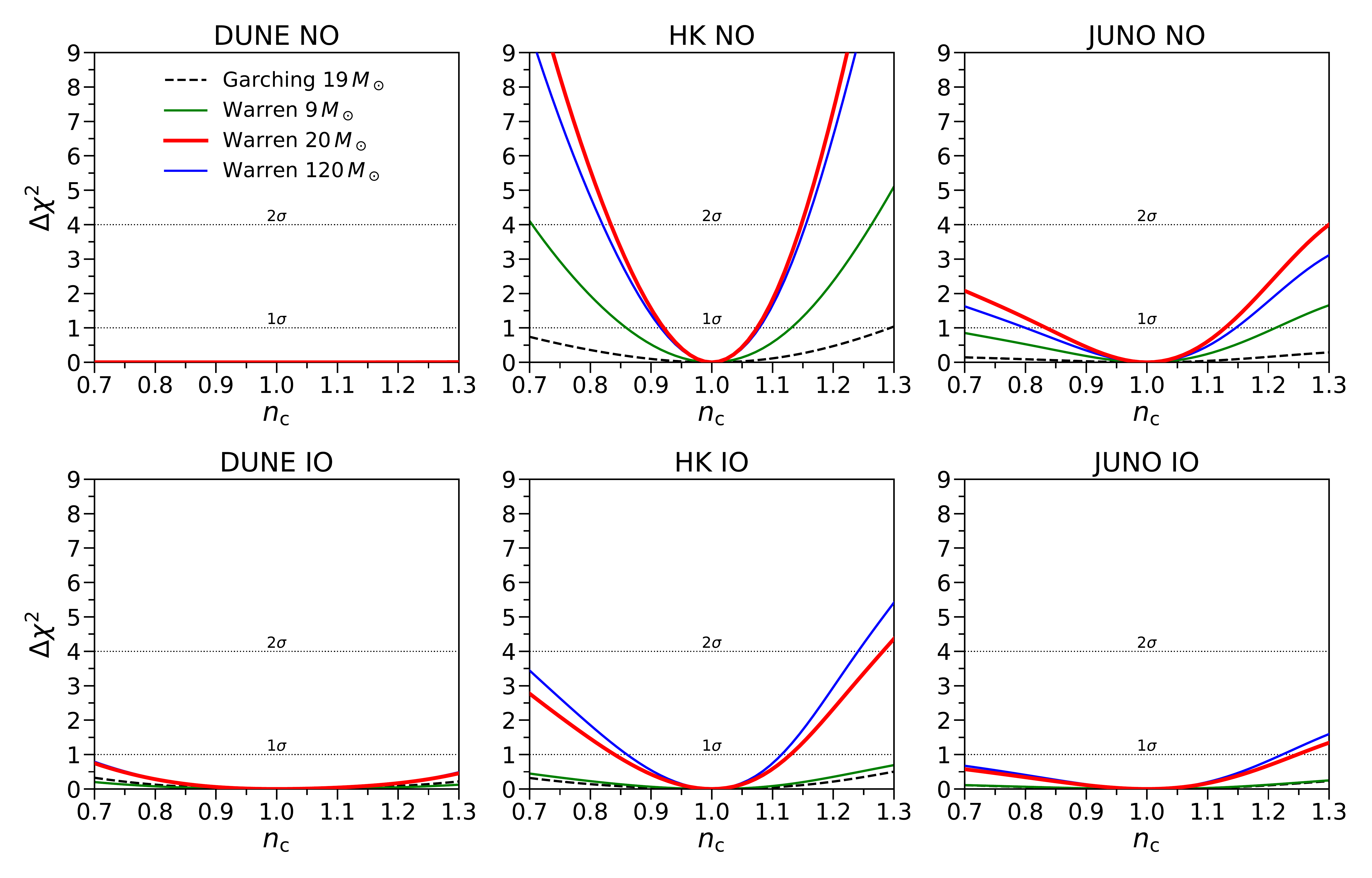}
\caption{\textit{\textbf{Dependence on SN neutrino spectra.}} Log-likelihood-ratio, $\Delta \chi^2$, as a function of the normalization parameter $n_c$ ($n_c = 1$ corresponds to the PREM profile) for various SN neutrino spectra: \texttt{Warren9} (green solid lines), \texttt{Warren20} (red solid lines), \texttt{Warren120} (blue solid lines)~\cite{Warren:2019lgb} and \texttt{Garching19} (black dashed lines)~\cite{Bollig:2020phc}. Results are shown for NO (top panels) and IO (bottom panels), for DUNE (left panels), HK (middle panels) and JUNO (right panels). We take a SN-Earth distance of $10$~kpc and the SN burst to occur on the opposite side of the detector (i.e., $c_z = - 1$).} 
\label{fig:chi2_models}
\end{center}
\end{figure}

We first study the capabilities of the three forthcoming neutrino detectors to determine the Earth density profile, assuming the SN neutrino spectra to be known, the SN burst to occur at 10~kpc and for neutrinos crossing the entire Earth ($c_z = - 1$). In general, this orientation corresponds to the most optimistic scenario, since the path of SN neutrinos inside the Earth is the longest one and thus, matter effects across the entire Earth would affect the neutrino propagation. 

The results are illustrated in Fig.~\ref{fig:chi2_models}, where the log-likelihood-ratio, $\Delta \chi^2$, is depicted as a function of the normalization parameter $n_c$ (recall that $n_c = 1$ recovers the PREM profile). We show results for the set of spectra shown in Fig.~\ref{fig:fluxes0}, for three progenitor masses ($9~M_\odot$, $20~M_\odot$ and $120~M_\odot$) obtained with the \texttt{Warren} simulations~\cite{Warren:2019lgb} and for a progenitor star with $18.88~M_\odot$ obtained with the \texttt{Garching19} simulations~\cite{Bollig:2020phc} for both the NO (top panels) and IO (bottom panels) cases. In the case of DUNE (left panels), the main detection channel probes the electron neutrino flux ($\nu_e\mathrm{Ar}-$CC), and therefore the best results, albeit very modest, are obtained for the IO case (bottom-left panel). For HK (middle panels) and JUNO (right panels), the main detection channel probes the electron antineutrino flux (IBD), being the best results obtained for NO (top-middle and top-right panels). The most sensitive detector to the Earth's density profile will be HK, due mainly to its larger mass, whereas DUNE will barely be able to provide meaningful constraints, due to its poor neutrino energy resolution at the relevant energies. Indeed, even for IO, the best sensitivity to matter effects would likely be reached by HK (bottom-middle panel), via subdominant $\nu_e-$CC interactions with oxygen, rather than by DUNE via the main $\nu_e\mathrm{Ar}-$CC detection channel. The projection for JUNO, with its smallest mass but with its superb energy resolution, is not as good as that of HK, but its sensitivity does not scale with the number of targets with respect to HK. 

Notice that the most optimistic results are obtained for the \texttt{Warren20} and \texttt{Warren120} spectra~\cite{Warren:2019lgb}, whereas the most pessimistic ones  are obtained for the \texttt{Garching19}~\cite{Bollig:2020phc} and  \texttt{Warren9}~\cite{Warren:2019lgb} SN neutrino spectra simulations. With any of the considered SN neutrino spectra, it will be challenging for DUNE to obtain an accuracy on the density profile (parameterized here by $n_c$) better than $\sim 30\%$, even at $1\sigma$~CL. On  the other hand, for the \texttt{Warren20} spectra~\cite{Warren:2019lgb}, HK could be able to establish $n_c = 1.00^{+0.07}_{-0.08}$ (NO) and $n_c = 1.00^{+0.13}_{-0.16}$ (IO) at $1\sigma$~CL, and JUNO may reach $n_c = 1.00^{+0.13}_{-0.17}$ (NO) at $1\sigma$~CL. Note that for IO, the \texttt{Warren120} spectra would render slightly more optimistic results.\footnote{In all cases, however, uncertainties comparable to geophysics ones, mainly from normal modes, would require a SN burst closer to us.} Obtaining any meaningful information about the Earth's density profile with the \texttt{Warren9} or \texttt{Garching19} spectra will be extremely  challenging, regardless of the underlying neutrino mass ordering. Indeed, relative differences among the results with different spectra can be quite substantial. They are mainly determined by the differences in the SN neutrino and antineutrino spectra at production. The largest the difference among the $\nu_e$ ($\bar{\nu}_e$) and  the $\nu_x$ ($\bar{\nu}_x$) spectra along the tail, the higher the sensitivity to matter effects after neutrinos traverse the Earth. Since the $\bar{\nu}_e$ and $\bar{\nu}_x$ spectra of the \texttt{Garching19} simulation are similar in the relevant energy range, matter effects for NO would get very diluted at detectors like HK and JUNO. 

We have also explicitly checked that these four different SN neutrino spectra could be distinguished from each other with a high confidence level, even when allowing for their overall normalizations to vary freely. This is expected, since with much smaller statistics than what we consider here, very precise model discrimination would be possible~\cite{Hyper-Kamiokande:2021frf}. Finally, we have also checked the impact of using time information by splitting the fluxes into three time bins: neutronization, accretion and cooling phases. In general, the main contribution comes from the cooling phase, but we do not find significant changes in sensitivity when performing time binning.

\subsection{Dependence on the energy resolution}
\label{sec:energyresolution}

The importance of energy resolution to observe Earth matter effects in SN neutrinos has been already discussed in the literature~\cite{Dighe:2003jg, Dighe:2003vm, Akhmedov:2005yt, Borriello:2012zc}. In mass-to-flavor transitions of MeV neutrinos in the Earth, attenuation of matter effects occurs at the detector due to the finite accuracy in reconstructing the neutrino energy~\cite{Ioannisian:2004jk, Ioannisian:2004vv, Ioannisian:2017chl, Bakhti:2020tcj}.\footnote{The term \textit{attenuation} could be misleading. It is related to the attenuation (or rather the wash out or weakening) of the matter effects from distant structures, and not to the attenuation of the incoming flux.} As a result, matter effects due to remote structures are suppressed at the detector. The extent of this wash-out effect is determined by both the oscillation length and the energy resolution of the experiment. The better the energy resolution, the more remote structures can be probed. Therefore, matter effects at distances beyond the attenuation length (or wash-out length) cannot be fully resolved at a given detector, since oscillations get averaged out.\footnote{This occurs at first order in the parameter $\epsilon$, Eq.~(\ref{eq:epsilon}). The attenuation effect is not present at order $\epsilon^2$~\cite{Ioannisian:2017chl}.} Such a limiting distance can be defined as~\cite{Ioannisian:2004jk, Ioannisian:2017chl}\footnote{In principle, the oscillation length in matter must be used. At the energies relevant for SN neutrinos, it is similar to that in vacuum. Note that even at the maximum of the regeneration factor, $\ell_0 = \ell_\oplus \, 2 \, \sin\theta_{12}$ (neutrinos) and $\ell_0 = \ell_\oplus \, 2 \, \cos\theta_{12}$ (antineutrinos), and therefore, both lengths are similar.} 
\begin{equation}
\label{eq:attlength}
\lambda_{\rm att} \equiv \ell_0 \, \left(\frac{E_\nu}{\pi \, \sigma_E}\right) = 4209~\mathrm{km} \, \left(\frac{E_\nu}{40~\mathrm{MeV}}\right) \, \left(\frac{7.5 \times 10^{-5}~\mathrm{eV}^2}{\Delta m^2_{21}}\right) \, \left(\frac{0.1}{\sigma_E/E_\nu}\right)  ~.
\end{equation}

Indeed, the distinct shape of the $\Delta \chi^2$ results with respect to varying $n_c$ shown in Fig.~\ref{fig:chi2_models} for DUNE (IO) as compared to HK and JUNO (NO) can also be partly explained by the different energy resolutions. The DUNE detector will have the poorest neutrino energy resolution at tens of MeV ($\sim 20\%$). Therefore, DUNE will be the least sensitive detector to matter effects in the Earth's core. For $n_c > 1$, the mantle density is smaller than in the PREM profile ($n_m < 1$). As DUNE is mostly sensitive to structures closer to the detector (the mantle), this implies that for $n_c > 1$, matter effects (in the mantle) shift to higher energies and thus, the sensitivity to the Earth's density profile is reduced. Yet, for the highest energies of the SN neutrino spectra, the wash-out effect is less important and some sensitivity to the core density remains. On the other hand, the energy resolution in HK and JUNO will be much better, and consequently the sensitivity to the core matter effect will be much less suppressed. For $n_c > 1$, at these detectors matter effects shift to lower energies. The behavior of the $\Delta \chi^2$ results with respect to $n_c$ is, thus, the opposite of DUNE. The relative performance between HK and JUNO, whose main detection channel is the same (IBD), can be understood in terms of their different size and their different energy resolution. Notice, however, that the effect of the intrinsic width of the positron energy distribution has a strong impact on JUNO. Its positron energy resolution is assumed to be better than $1\%$ for $E_\nu > 20$~MeV, however at those energies the width of the positron energy distribution is $> 2\%$: it is this latter figure what sets the largest distance that can be probed. Coincidentally for IBD, this intrinsic uncertainty on the neutrino energy ($\sim E_\nu/m_p$) implies a maximum attenuation length of the size of the entire Earth, $\left(\lambda_\mathrm{att}\right)_\mathrm{max} \sim 10^4$~km, regardless of the precise value of the neutrino energy.

\begin{figure}[t]
\begin{center}
\includegraphics[width=\columnwidth]{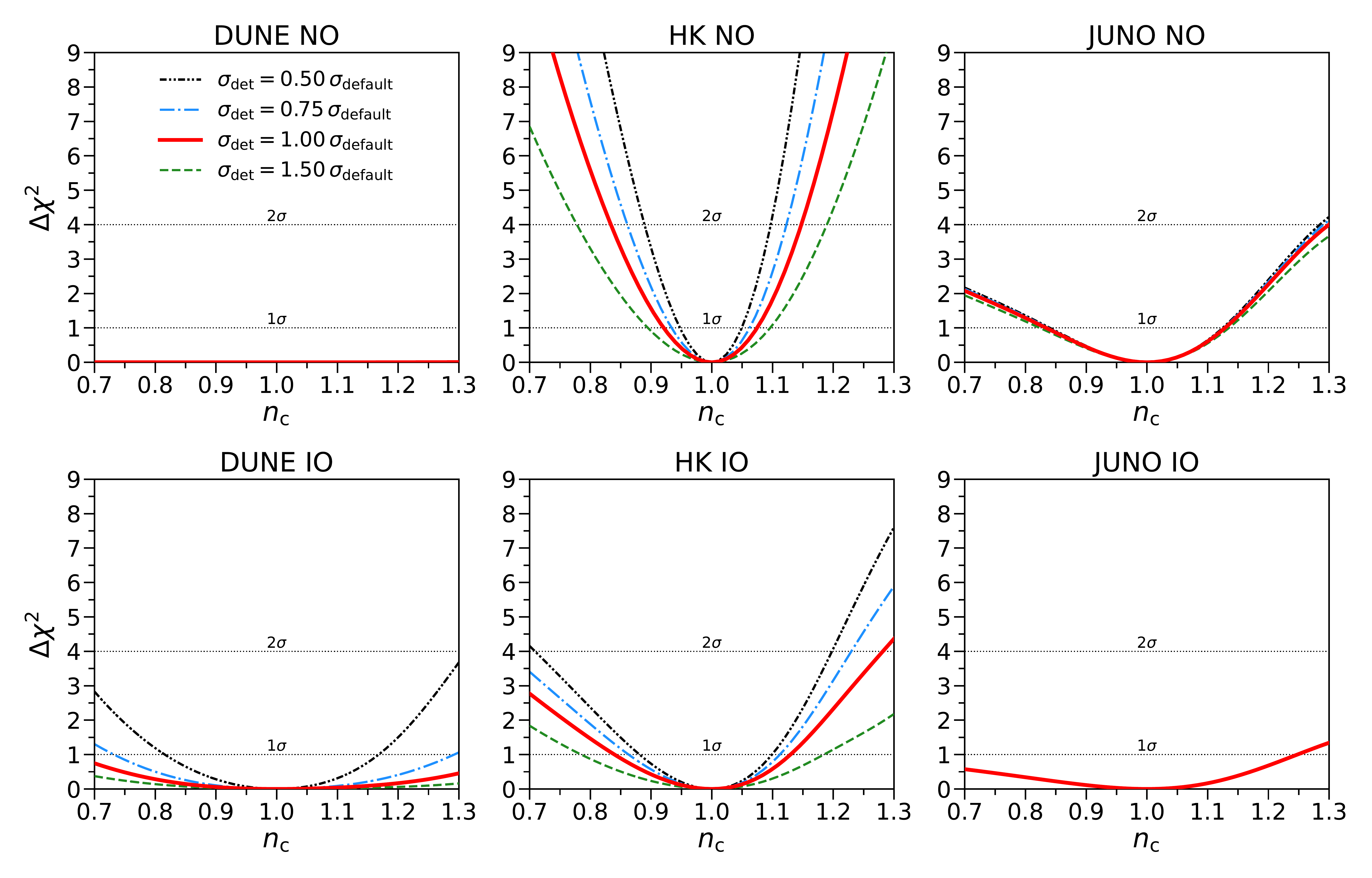}
\caption{\textit{\textbf{Dependence on energy resolution.}} Log-likelihood-ratio, $\Delta \chi^2$, as a function of the normalization parameter $n_c$ ($n_c = 1$ corresponds to the PREM profile) for different energy resolutions: $\sigma_\mathrm{det} = f \, \sigma_\mathrm{default}$, where $f =$ 0.5 (black dot-dot-dashed lines), 0.75 (blue dot-dashed lines), 1 (red solid lines) and 1.5 (green dashed lines). Results are shown for NO (top panels) and IO (bottom panels), for DUNE (left panels), HK (middle panels) and JUNO (right panels), using the \texttt{Warren20} SN neutrino spectra~\cite{Warren:2019lgb}. We take a SN-Earth distance of $10$~kpc and the SN burst to occur on the opposite side of the detector (i.e., $c_z = - 1$).} 
\label{fig:chi2_res}
\end{center}
\end{figure}

Figure~\ref{fig:chi2_res} depicts the effect of the energy resolution on the determination of the Earth's density profile, using our fiducial \texttt{Warren20} SN neutrino spectra~\cite{Warren:2019lgb} and $c_z = -1$. We show the log-likelihood-ratio, $\Delta \chi^2$, as a function of $n_c$, varying the default energy resolution of each detector by multiplying their assumed default value by an overall factor. If the energy resolution is improved by a factor of two, the sensitivity to the density profile of DUNE and HK would significantly improve, but instead it would barely change in the case of JUNO. For $\sigma_\mathrm{det} = 0.5 \, \sigma_\mathrm{default}$, at $1\sigma$ CL, $n_c = 1.00^{+0.17}_{-0.18}$ (DUNE, IO), $n_c = 1.00 \pm 0.05$ (HK, NO), $n_c = 1.00^{+0.10}_{-0.12}$ (HK, IO) and $n_c = 1.00^{+0.12}_{-0.16}$ (JUNO, NO). This can be understood from the comparison of the bin size with the visible energy resolution and with the width of the positron distribution in IBD. We are using the very same energy binning for all three detectors, $\Delta E_\mathrm{rec} = 2$~MeV. In the case of DUNE and HK, it is smaller than the energy resolution at tens of MeV, which in turn is wider than the spread in the positron energy in IBD. Thus, improving the energy resolution makes a difference in resolving features in the neutrino spectra and hence, in resolving matter effects. Even better energy resolution than what we consider in Fig.~\ref{fig:chi2_res} would further improve HK sensitivity. In the very same way, worse energy resolution in DUNE and HK would result into a non-negligible loss of sensitivity to the density profile. On the contrary, for the JUNO detector not only the bin size but also the width of the positron distribution are larger than the assumed $\sim 1\%$ visible energy resolution at the relevant energies. Thus, the results would change very little by slightly improving the energy resolution and more importantly, a slightly worse energy reconstruction would not result in a loss of sensitivity either. We have also checked that reducing the bin size does not significantly improve the results. This is mainly due to the fact that the intrinsic spread of the positron (in IBD) at the relevant energies (tens of MeV) is comparable or larger than $\Delta E_\mathrm{rec} = 2$~MeV.

\subsection{Dependence on the SN direction with respect to the detector}

\begin{figure}[t]
\begin{center}
\includegraphics[width=\columnwidth]{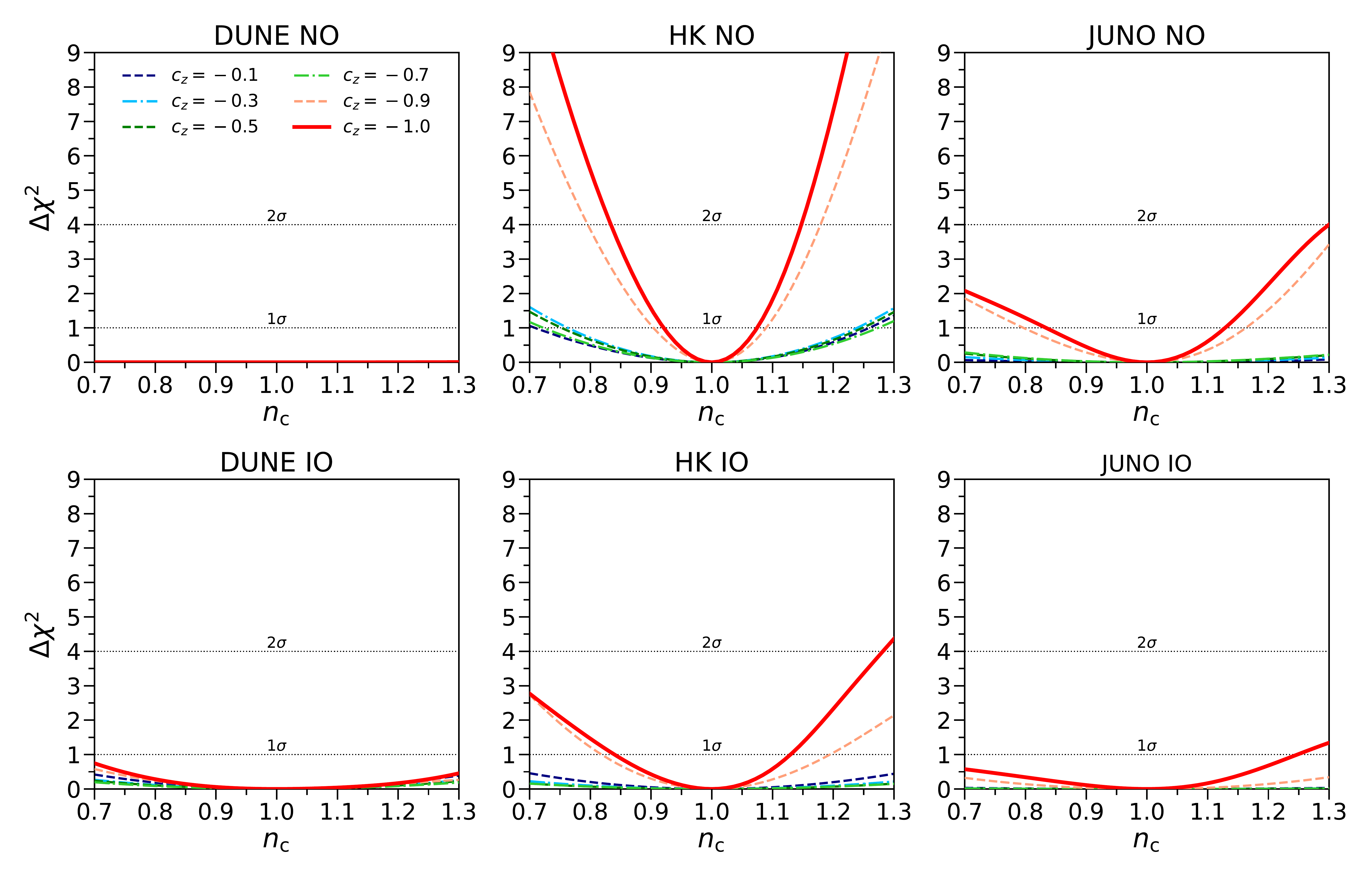}
\caption{\textit{\textbf{Dependence on SN direction.}} Log-likelihood-ratio, $\Delta \chi^2$, as a function of the normalization parameter $n_c$ ($n_c = 1$ corresponds to the PREM profile) for various trajectories through the Earth: $c_z = -1$ (red solid lines), $-0.9$ (dashed orange lines), $-0.7$ (light green dot-dashed lines), $-0.5$ (dark green dashed lines), $-0.3$ (light blue dot-dashed lines) and $-0.1$ (dark blue dashed lines). Results are shown for NO (top panels) and IO (bottom panels), for DUNE (left panels), HK (middle panels) and JUNO (right panels), using the \texttt{Warren20} SN neutrino spectra~\cite{Warren:2019lgb}. We take a SN-Earth distance of $10$~kpc.} 
\label{fig:chi2_cos}
\end{center}
\end{figure}

All the results previously presented have been restricted to the case of SN neutrinos traveling through the entire Earth ($c_z = -1$). In the following, we shall study how the sensitivity to the density profile gets modified for trajectories that partially cross the Earth. For zenith angles such that $c_z > - 0.838$, neutrinos only traverse the mantle. In Fig.~\ref{fig:chi2_cos} we show the log-likelihood-ratio, $\Delta \chi^2$, as a function of $n_c$, for different trajectories using the \texttt{Warren20} SN neutrino spectra~\cite{Warren:2019lgb}. This set of SN spectra turns out to be the most optimistic case among the four we consider for $c_z = -1$ and NO, although this is not the case for other possible trajectories (or for IO). As a general feature, the sensitivity to structures of the density profile of size $\sim \ell_\oplus/2$ is enhanced, as expected~\cite{Bakhti:2020tcj}. This implies a local maximum in the $\Delta \chi^2$ results for trajectories with $c_z \sim (-0.07,-0.1)$, corresponding to a path length $\sim \ell_\oplus/2$ for energies $E_\nu \sim (50-70)$~MeV, which are the energies most sensitive to Earth matter effects. In general, though, this is not the global maximum as a function of $c_z$. Trajectories crossing the Earth core would typically result in the highest sensitivity to the density profile, although the shape of neutrino spectra and the detector's energy resolution can slightly alter this conclusion. Indeed, the sensitivity to matter effects for (anti)neutrinos crossing only the mantle is very reduced in all cases, but it would be enhanced if the Earth's core is traversed. This is an effect of the energy resolution and the related sensitivity to more remote distances. Improving energy reconstruction significantly affects the sensitivity to matter effects for core-crossing neutrinos, but has a smaller impact for neutrinos crossing only the mantle. This can be noticed from the DUNE results (left panels). For $n_c > 1$, the trajectory with $c_z = -1$ is similar to other trajectories through the core and even to some trajectories only through the mantle. With a better energy resolution, though, the best sensitivity would always be for trajectories crossing the core. Indeed, this can be seen from the results for HK (middle panels) and JUNO (right panels).

\subsection{Dependence on $\Delta m_{21}^2$ and $\theta_{12}$}

\begin{figure}[t]
\begin{center}
\includegraphics[width=\columnwidth]{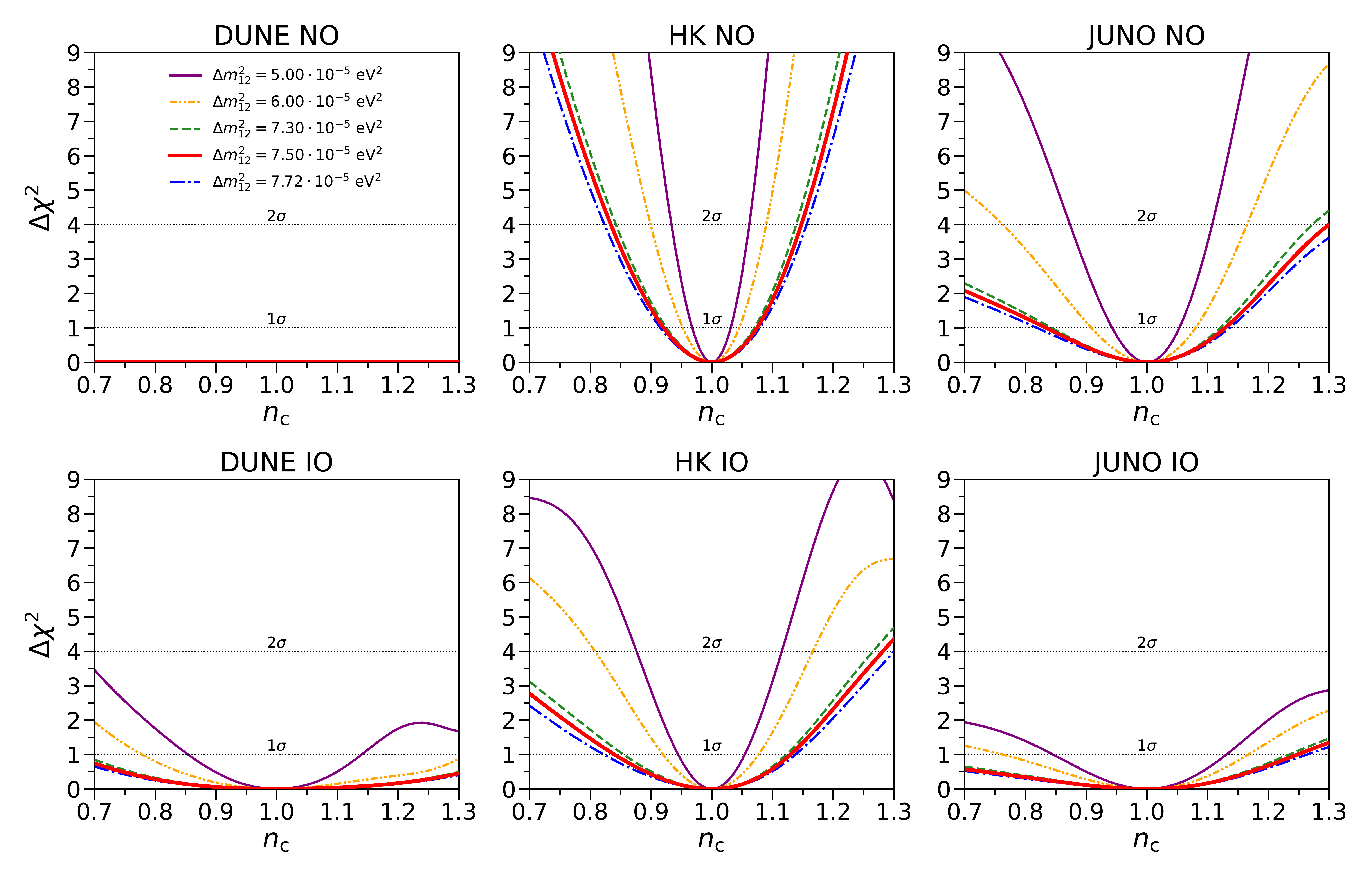}
\caption{\textit{\textbf{Dependence on $\boldsymbol{\Delta m_{21}^2}$.}} Log-likelihood-ratio, $\Delta \chi^2$, as a function of the normalization parameter $n_c$ ($n_c = 1$ corresponds to the PREM profile) for five values of $\Delta m_{21}^2$: the current global best fit, $7.5 \times 10^{-5}~\mathrm{eV}^2$ (red solid lines), the limits of the $\pm 1\sigma$~CL interval, $7.30 \times 10^{-5}~\mathrm{eV}^2$ (green dashed lines) and $7.72 \times 10^{-5}~\mathrm{eV}^2$ (blue dot-dashed lines)~\cite{deSalas:2020pgw}, $5 \times 10^{-5}~\mathrm{eV}^2$ (violet solid lines) and $6 \times 10^{-5}~\mathrm{eV}^2$ (orange dot-dot-dashed lines). Results are shown for NO (top panels) and IO (bottom panels), for DUNE (left panels), HK (middle panels) and JUNO (right panels), using the \texttt{Warren20} SN neutrino spectra~\cite{Warren:2019lgb}. We take a SN-Earth distance of $10$~kpc and the SN burst to occur on the opposite side of the detector (i.e., $c_z = - 1$).}
\vspace{-3mm}
\label{fig:chi2_dm12}
\end{center}
\end{figure}

The impact of the solar mass-squared difference, $\Delta m^2_{21}$, on Earth matter effects for SN neutrinos was studied long ago~\cite{Lunardini:2000sw, Lunardini:2001pb, Takahashi:2001dc, Dighe:2003jg}, although mainly focusing on relatively small values, $\Delta m^2_{21} \lesssim 6 \times 10^{-5}~\mathrm{eV}^2$, compared to the current global best fit~\cite{deSalas:2020pgw, Esteban:2020cvm, Capozzi:2017ipn}. This is particularly important, since for SN neutrinos, Earth matter effects are more pronounced the smaller $\Delta m_{21}^2$ is. In turn, this implies longer oscillation lengths, but also longer attenuation lengths, which results in enhanced sensitivity to remote structures. This dependence is illustrated in Fig.~\ref{fig:chi2_dm12}, where we show the log-likelihood-ratio, $\Delta \chi^2$, as a function of $n_c$, assuming $c_z = -1$ and the \texttt{Warren20} SN neutrino spectra~\cite{Warren:2019lgb}. We consider five values of $\Delta m^2_{21}$, fixing the remaining neutrino oscillation parameters to their current best-fit values. We depict results for the current global best fit of $\Delta m^2_{21}$, together with those obtained for its $\pm 1\sigma$~CL allowed limits~\cite{deSalas:2020pgw}. We also present the results for (rounded) values of $\Delta m^2_{21}$ close to the best fit with pre-2020 SK data~\cite{Super-Kamiokande:2016yck, Nakano:2020fgc} (which is also the value used in the earlier tomography study with SN neutrinos~\cite{Lindner:2002wm}), $\Delta m_{21}^2 = 5 \times 10^{-5}~\mathrm{eV}^2$, and to the current best fit using solar neutrino data after different improvements in the SK analysis~\cite{Solar2022}, $\Delta m_{21}^2 = 6 \times 10^{-5}~\mathrm{eV}^2$. Solar neutrino data alone favor values smaller than the current global best fit, which is dominated by the KamLAND result~\cite{KamLAND:2010fvi, KamLAND:2013rgu}. After the latest SK improvements, this tension is reduced to $\sim 1.5\sigma$~\cite{Solar2022}. Whereas the uncertainty of the global fit has a mild effect on our results, much smaller values of $\Delta m_{21}^2$, as preferred by solar data alone, would have a very strong impact. The resulting HK constraints (for NO) at $\sim 3\sigma$~CL with $\Delta m_{21}^2$ from solar data would be similar to those obtained at $\sim 2\sigma$~CL with the current global best fit. For $\Delta m_{21}^2 = 5 \times 10^{-5}~\mathrm{eV}^2$, the HK constraints (for NO) at $\sim 3\sigma$~CL would be similar to those obtained at $\sim 1\sigma$~CL with the $\Delta m_{21}^2$ global best fit (i.e., $\lesssim 10\%$ on $n_c$). This is slightly less optimistic than the results of an early study of the Earth's density profile with SN neutrinos, which concluded that the average core density could be determined at the percent level with a significance of 2$\sigma$~\cite{Lindner:2002wm}.

 \begin{figure}[t]
\begin{center}
\includegraphics[width=\columnwidth]{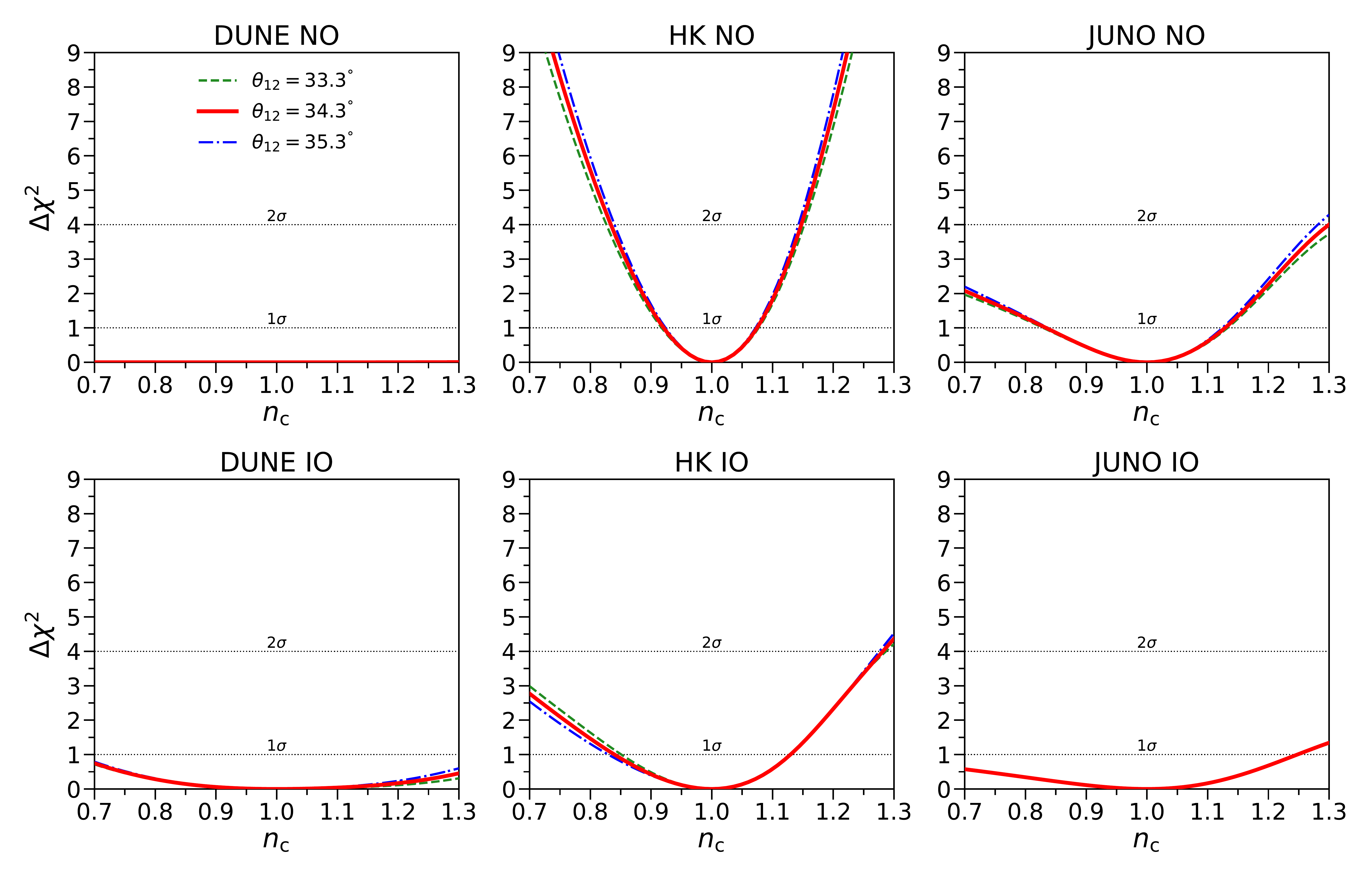}
\caption{\textit{\textbf{Dependence on $\boldsymbol{\theta_{12}}$.}} Same as Fig.~\ref{fig:chi2_dm12}, but for three values of $\theta_{12}$: the current best fit, $34.3^\circ$ (red solid lines), and its $\pm 1\sigma$~CL limits, $33.3^\circ$ (green dashed lines) and $35.3^\circ$ (blue dot-dashed lines)~\cite{deSalas:2020pgw}.}
\label{fig:chi2_theta12}
\end{center}
\end{figure}

The next important parameter is $\theta_{12}$. At first order, it enters the matter-dependent term in the transition probabilities with a linear dependence on $\sin^2 2\theta_{12}$, see Eq.~(\ref{eq:pmatter}). For both neutrinos and antineutrinos, the larger $\theta_{12}$ is, the more important matter effects are.  In Fig.~\ref{fig:chi2_theta12} we show the $\Delta \chi^2$ results, as a function of $n_c$, for the global best fit of $\theta_{12}$ and for its $\pm 1\sigma$~CL limits. For the rest of the inputs, we make the same assumptions as in Fig.~\ref{fig:chi2_dm12}. The impact of the uncertainty on $\theta_{12}$ is slightly smaller than that of $\Delta m_{21}^2$, and in general it is quite mild.

\subsection{Sensitivity to the electron fraction in the core} 

Coherent matter effects in neutrino propagation are proportional to the electron density. Thus, it is the product of the total density times the electron fraction what determines the matter potential (i.e., $Y_e \, \rho$), see Eq.~(\ref{eq:epsilon}). So far, we have kept the electron fraction fixed to its PREM values. Yet, it is interesting to estimate to which extent the chemical composition of Earth could be determined with SN neutrinos. Indeed, this has critical importance in geophysics and even a small amount of light elements in the core would have profound consequences~\cite{Hirose:2021}.   

We have obtained the sensitivity of future neutrino detectors to the electron fraction in the core by freely varying $Y_{e,m}$ and $Y_{e,c}$, fixing the density profile to the PREM profile ($n_m = n_c = 1$). This is equivalent to not imposing the constraint on the total mass of the Earth and fixing $Y_{e,m}$ and $Y_{e,c}$. With respect to the results shown so far, the sensitivity of the three detectors gets degraded, the least affected being JUNO. This is expected, since its superb energy resolution would allow it to retain more of its sensitivity to more remote structures and thus, to the electron fraction in the core. In any case, with the inputs and parameters assumed in this study, it will be very challenging to determine $Y_{e,c}$ (or $Y_{e,m}$) with an accuracy better than $\sim 10\%$ at $1\sigma$~CL with HK, and it would be worse with JUNO or DUNE. All in all, a core composition of pure iron/nickel differs from one with a $\sim 1\%$ weight fraction of hydrogen by $\Delta Y_{e, c} \sim 0.005$. Hence, a $\sim 1\%$ precision would be required to distinguish these two extreme compositions. Probably, the only way to achieve such a precision with SN neutrinos would be a SN explosion much closer than 10~kpc.

\section{Summary and conclusions} 
\label{sec:conclusions}

Determining the internal structure of the Earth is crucial to understand its dynamics and thermal history. Traditional methods to determine its density profile mainly rely on the study of seismic waves, along with constraints from gravitational measurements. A completely different method, both conceptually and methodologically, which unlike traditional ones, relies purely on weak interactions, is neutrino tomography. As a matter of fact, although the idea of using neutrinos for Earth absorption tomography is five decades old~\cite{Placci:1973, Volkova:1974xa}, it was only a few years ago when the first Earth absorption tomography with available atmospheric neutrino data was performed~\cite{Donini:2018tsg, Salvado:2019hfn}. On the other hand, the idea of Earth neutrino oscillation tomography dates back to the mid 1980's~\cite{Ermilova:1986ph, Nicolaidis:1987fe, Ermilova:1988pw, Nicolaidis:1990jm}, although neutrino data, from different sources and energy ranges, are not abundant and precise enough yet to perform such a study. 

In this work, we have revisited the possibility to perform Earth (oscillation) tomography with SN neutrinos at future neutrino detectors (DUNE, HK and JUNO), operating at planned facilities. We have considered up-to-date calculations of SN neutrino spectra and the most recent neutrino mixing parameters. These updates are of particular importance in the case of the value of the solar mass-squared difference, $\Delta m_{21}^2$. Earth matter effects on SN neutrinos, with energies of tens of MeV, are controlled by the term $2 E_\nu V/\Delta m^2_{21}$~\cite{Minakata:1987fj, Smirnov:1993ku, Dighe:1999bi, Lunardini:2000sw, Takahashi:2000it, Lunardini:2001pb, Takahashi:2001dc, Fogli:2001pm, Lunardini:2003eh, Dighe:2003jg, Dighe:2003vm, Dasgupta:2008my, Guo:2008mma, Scholberg:2009jr, Borriello:2012zc}. An earlier Earth tomography study with SN neutrinos considered $\Delta m_{21}^2 = 5 \times 10^{-5}~\mathrm{eV}^2$~\cite{Lindner:2002wm}, which is significantly smaller than the current global best fit, $\Delta m_{21}^2 = \left(7.5 ^{+ 0.22}_{-0.20}\right) \times 10^{-5}~\mathrm{eV}^2$~\cite{deSalas:2020pgw}. Thus, the results of that analysis were likely too optimistic. Furthermore, new detectors have been proposed in the last decades, so an updated analysis was timely.

After introducing in Sec.~\ref{sec:fluxes} the set of SN neutrino spectra we consider (Fig.~\ref{fig:fluxes0}) and discussing the main features of Earth matter effects on the fluxes arriving to Earth (Fig.~\ref{fig:prob}), in Sec.~\ref{sec:events}, we describe the main detection characteristics of the three future neutrino detectors we study in this work: DUNE, HK and JUNO, depicting the expected event distributions (Figs.~\ref{fig:totalrate} and~\ref{fig:reldevrate}). We argue that, in order to properly compute the expected signal event spectra, differential cross sections must be used. The details of the analysis and the main results of the paper are provided in Sec.~\ref{sec:analysis}, where we describe the dependence of the sensitivity to the Earth density profile on several factors (Figs.~\ref{fig:chi2_models}--\ref{fig:chi2_theta12}).

Matter effects driven by $\Delta m_{21}^2$ are maximal for $E_\nu \sim (40 - 100)$~MeV (see Eq.~(\ref{eq:epsilon})), depending on the neutrino trajectory through the Earth. Thus, they are more important along the high-energy tail of the SN neutrino spectra. Nevertheless, the exponential decrease of the flux results in low statistics at the highest energies, and therefore, most of the sensitivity to the density profile of the Earth lies in the interval $E_\nu \sim (40 - 80)$~MeV. In this work, we have studied these effects at DUNE, HK and JUNO, assuming a SN-Earth distance of 10~kpc. Assuming adiabatic propagation in the interior of the SN, Earth matter effects mostly occur for $\nu_e$ in the case of IO and for $\bar{\nu}_e$ in the case of NO. Thus, for DUNE, IO is the most optimistic case (its main detection channel is $\nu_e$Ar$-$CC interactions), whereas for HK and JUNO it is NO (their main detection channel is IBD). We have studied the dependence on different SN neutrino spectra (Fig.~\ref{fig:chi2_models}). We have also made special emphasis on the impact of neutrino energy resolution (Fig.~\ref{fig:chi2_res}) and its connection to resolving remote structures. The dependence of our results on the SN direction with respect to the detector has also been discussed (Fig.~\ref{fig:chi2_cos}) and we conclude that, for the current best-fit $\Delta m_{21}^2$, only for core-crossing neutrinos significant sensitivity to the Earth density profile may be obtained. Given that matter effects grow for smaller $\Delta m_{21}^2$ and for larger $\theta_{12}$, it is important to evaluate the effect of varying these two parameters (Figs.~\ref{fig:chi2_dm12} and~\ref{fig:chi2_theta12}). This also allows us to compare our results to previous ones that used values of $\Delta m_{21}^2$ smaller than the current global best fit. Indeed, turning the question around, Earth matter effects could be exploited to determine these parameters, which would be presented elsewhere~\cite{Hajjar:2023xae}. Finally, we have also discussed the potential sensitivity to the electron fraction in the core, of critical importance in geophysics. We have shown that constraints on the density and the electron fraction at the level of $\lesssim 10\%$ at $1\sigma$~CL could be achieved at HK if SN neutrinos cross the Earth's core. In this regard, a closer SN neutrino burst would be required to obtain uncertainties comparable to those from geophysics. 

In summary, we have shown that future detectors, such as HK and JUNO, could determine the average Earth's core density within $\lesssim 10\%$ at $1\sigma$~CL with galactic SN neutrinos (at 10~kpc). To achieve a similar sensitivity, the neutrino energy reconstruction capabilities of the DUNE detector should be improved. All in all, even if less optimistic than earlier results, we have shown that a future SN burst could aid in future neutrino Earth tomography studies, and be competitive with, and complementary to other analyses considering other neutrino sources and energy ranges. 

\vspace{-2mm}
\begin{acknowledgments}
\vspace{-1mm}

We thank L.~Álvarez-Ruso, N.~Jachowicz and T. Suzuki for discussions and J.~Salvado for help with the \texttt{nuSQuIDS} code. This work has been supported by the Spanish MCIN/AEI/10.13039/501100011033 grants PID2020-113644GB-I00 (RH and OM) and PID2020-113334GB-I00 (SPR) and by the European ITN project HIDDeN (H2020-MSCA-ITN-2019/860881-HIDDeN). RH is supported by the Spanish grant FPU19/03348 of MU. The authors also acknowledge support from the Generalitat Valenciana grants PROMETEO/2019/083 and CIPROM/2022/69 (RH and OM) and CIPROM/2022/36 (SPR). SPR was also partially supported by the Portuguese FCT (CERN/FIS-PAR/0004/2019 and CERN/FIS-PAR/0019/2021). 
\end{acknowledgments}
\vspace{-1mm}

\bibliography{biblio}
\bibliographystyle{apsrev4-1}
\end{document}